\documentclass[10pt]{llncs}

\mainmatter
\makeatletter
\@ifundefined{lhs2tex.lhs2tex.sty.read}%
  {\@namedef{lhs2tex.lhs2tex.sty.read}{}%
   \newcommand\SkipToFmtEnd{}%
   \newcommand\EndFmtInput{}%
   \long\def\SkipToFmtEnd#1\EndFmtInput{}%
  }\SkipToFmtEnd

\newcommand\ReadOnlyOnce[1]{\@ifundefined{#1}{\@namedef{#1}{}}\SkipToFmtEnd}
\usepackage{amstext}
\usepackage{amssymb}
\usepackage{stmaryrd}
\DeclareFontFamily{OT1}{cmtex}{}
\DeclareFontShape{OT1}{cmtex}{m}{n}
  {<5><6><7><8>cmtex8
   <9>cmtex9
   <10><10.95><12><14.4><17.28><20.74><24.88>cmtex10}{}
\DeclareFontShape{OT1}{cmtex}{m}{it}
  {<-> ssub * cmtt/m/it}{}

\DeclareFontShape{OT1}{cmtt}{bx}{n}
  {<5><6><7><8>cmtt8
   <9>cmbtt9
   <10><10.95><12><14.4><17.28><20.74><24.88>cmbtt10}{}
\DeclareFontShape{OT1}{cmtex}{bx}{n}
  {<-> ssub * cmtt/bx/n}{}
\newcommand{\Conid}[1]{\mathit{#1}}
\newcommand{\Varid}[1]{\mathit{#1}}
\newcommand{\anonymous}{\kern0.06em \vbox{\hrule\@width.5em}}

\newcommand{\bind}{\mathbin{>\!\!\!>\mkern-6.7mu=}}
\renewcommand{\leq}{\leqslant}

\usepackage{polytable}

\@ifundefined{mathindent}%
  {\newdimen\mathindent\mathindent\leftmargini}%
  {}%

\def\resethooks{%
  \global\let\SaveRestoreHook\empty
  \global\let\ColumnHook\empty}
\newcommand*{\savecolumns}[1][default]%
  {\g@addto@macro\SaveRestoreHook{\savecolumns[#1]}}
\newcommand*{\restorecolumns}[1][default]%
  {\g@addto@macro\SaveRestoreHook{\restorecolumns[#1]}}
\newcommand*{\aligncolumn}[2]%
  {\g@addto@macro\ColumnHook{\column{#1}{#2}}}

\resethooks

\newcommand{\onelinecommentchars}{\quad-{}- }
\newcommand{\commentbeginchars}{\enskip\{-}
\newcommand{\commentendchars}{-\}\enskip}

\newcommand{\visiblecomments}{%
  \let\onelinecomment=\onelinecommentchars
  \let\commentbegin=\commentbeginchars
  \let\commentend=\commentendchars}

\newcommand{\invisiblecomments}{%
  \let\onelinecomment=\empty
  \let\commentbegin=\empty
  \let\commentend=\empty}

\visiblecomments

\newlength{\blanklineskip}
\newcommand{\hsindent}[1]{\quad}% default is fixed indentation
\let\hspre\empty
\let\hspost\empty

\EndFmtInput
\makeatother
\ReadOnlyOnce{polycode.fmt}%
\makeatletter

\newcommand{\hsnewpar}[1]%
  {{\parskip=0pt\parindent=0pt\par\vskip #1\noindent}}

\newcommand{\hscodestyle}{}

\newcommand{\sethscode}[1]%
  {\expandafter\let\expandafter\hscode\csname #1\endcsname
   \expandafter\let\expandafter\endhscode\csname end#1\endcsname}

  {\par\noindent
   \advance\leftskip\mathindent
   \hscodestyle
   \let\\=\@normalcr
   \let\hspre\(\let\hspost\)%
   \pboxed}%
  {\endpboxed\)%
   \par\noindent
   \ignorespacesafterend}

  {\hsnewpar\abovedisplayskip
   \advance\leftskip\mathindent
   \hscodestyle
   \let\hspre\(\let\hspost\)%
   \pboxed}%
  {\endpboxed%
   \hsnewpar\belowdisplayskip
   \ignorespacesafterend}

  {\hsnewpar\abovedisplayskip
   \advance\leftskip\mathindent
   \hscodestyle
   \let\\=\@normalcr
   \(\pboxed}%
  {\endpboxed\)%
   \hsnewpar\belowdisplayskip
   \ignorespacesafterend}

\newcommand{\plainhs}{\sethscode{plainhscode}}

\plainhs

  {\hsnewpar\abovedisplayskip
   \advance\leftskip\mathindent
   \hscodestyle
   \let\\=\@normalcr
   \(\parray}%
  {\endparray\)%
   \hsnewpar\belowdisplayskip
   \ignorespacesafterend}

  {\parray}{\endparray}

  {\(\parray}{\endparray\)}

\def\codeframewidth{\arrayrulewidth}
\RequirePackage{calc}

  {\parskip=\abovedisplayskip\par\noindent
   \hscodestyle
   \arrayrulewidth=\codeframewidth
   \tabular{@{}|p{\linewidth-2\arraycolsep-2\arrayrulewidth-2pt}|@{}}%
   \hline\framedhslinecorrect\\{-1.5ex}%
   \let\endoflinesave=\\
   \let\\=\@normalcr
   \(\pboxed}%
  {\endpboxed\)%
   \framedhslinecorrect\endoflinesave{.5ex}\hline
   \endtabular
   \parskip=\belowdisplayskip\par\noindent
   \ignorespacesafterend}

\newcommand{\framedhslinecorrect}[2]%
  {#1[#2]}

  {\(\def\column##1##2{}%
   \let\>\undefined\let\<\undefined\let\\\undefined
   \newcommand\>[1][]{}\newcommand\<[1][]{}\newcommand\\[1][]{}%
   \def\fromto##1##2##3{##3}%
   }{\) }%

  {\let\orighscode=\hscode
   \let\origendhscode=\endhscode
   \def\endhscode{\def\hscode{\endgroup\def\@currenvir{hscode}\\}\begingroup}
   \orighscode\def\hscode{\endgroup\def\@currenvir{hscode}}}%
  {\origendhscode
   \global\let\hscode=\orighscode
   \global\let\endhscode=\origendhscode}%

\makeatother
\EndFmtInput

\usepackage{multicol}
\usepackage{amsmath}
\usepackage{amssymb,latexsym}
\usepackage{amsmath}
\usepackage{verbatim}
\usepackage{graphicx}

\usepackage{wrapfig}
\usepackage{lipsum}

\usepackage{url}

\usepackage{xspace}
\usepackage[sharp]{easylist}

\usepackage{appendix}

\usepackage{tabularx}

\let\temp\phi
\let\phi\varphi
\let\varphi\temp

\usepackage{todonotes}
\usepackage{blkarray}

\usepackage{ltlfonts}
\usepackage{listings}
\usepackage{tabu}
\usepackage[english]{babel}
\usepackage{paralist}
\usepackage{multirow}
\usepackage{verbatimbox}
\usepackage{fancyvrb}

\definecolor{StriverBlue}{RGB}{42, 0, 255}
\definecolor{StriverGreen}{RGB}{0, 155, 0}
\definecolor{StriverRed}{RGB}{155, 0, 0}
\definecolor{StriverOrange}{RGB}{255, 183, 68}
\definecolor{StriverYellow}{RGB}{155, 155, 45}
\lstdefinelanguage{Striver}
{
    basicstyle=\ttfamily,
    keywordstyle=[1]\color{StriverYellow},
    keywordstyle=[2]\color{StriverBlue},
    keywordstyle=[3]\slshape,
    keywordstyle=[4]\color{StriverBlue},
    keywordstyle=[5]\color{StriverOrange},
    keywordstyle=[6]\color{StriverGreen},
    keywordstyle=[7]\color{StriverRed},
    otherkeywords = {
    <~,<<,>~,>>,:=,||,&&,+,-,==,!,!=, =>},
    morekeywords = [1]{<~,<<,>~,>>},
    morekeywords = [2]{shift,macro, :=, input,const, output, ticks, define},
    morekeywords = [3]{out, notick, true, false, cv},
    morekeywords = [4]{if, then, else},
    morekeywords = [5]{char, int, double, bool, unit, Time, Time_eps, void},
    morekeywords = [6]{U, delay, ||, &&, +, -, !, !=, ==, at, =>},
    morekeywords = [7]{let, in},
}

\lstdefinelanguage{LOLA}
{
    escapechar={\~}, 
    basicstyle=\ttfamily,
    otherkeywords = {macro,:=,const,input,output,true,false,define,outside,U,delay,||,&&,+,-,==,!,!=,if,let,in then,else,int,bool,unit,Time,num},
    morekeywords = [1]{},
    morekeywords = [2]{macro, :=, input, const, output,  define},
    morekeywords = [3]{true, false},
    morekeywords = [4]{if, then, else},
    morekeywords = [5]{int, bool, unit, num},
    morekeywords = [6]{U, delay, ||, &&, +, -, !, !=, ==, or, and, not, implies},
    morekeywords = [7]{let, in},
}

\newcommand{\hLola}{\textsc{hLola}\xspace}
\newcommand{\hlola}{\textsc{\hLola}\xspace}

\newcommand \hi[1]{}

\newcommand \sizeof[1] { \ensuremath{\mathbin{\#}\mathrm{1}} }
\newcommand \lola{\textsc{Lola}\xspace}

\newcommand \false {\mathit{false}}
\newcommand \true {\mathit{true}}

\newcommand{\IfThenElse}[3]{\ensuremath{\textup{\texttt{if}}\;#1\;\textup{\texttt{then}}\;#2\;\textup{\texttt{else}}\;#3}}

\renewcommand{\Or}{\mathrel{\vee}}

\newcommand{\Into}{\mathrel{\rightarrow}}

\newcommand{\Always}{\LTLsquare}
\newcommand{\Event}{\LTLdiamond} 

\newcommand{\LTLPrev}{\LTLcircleminus}

\newcommand{\HasAlwaysBeen}{\LTLsquareminus}
\newcommand{\Once}{\LTLdiamondminus}
\newcommand{\Since}{\mathbin{\mathcal{S}}}

\newcommand{\UNTIL}{\mathbin{\mathcal{U}}}

\newcommand{\BoolT}{\textit{Bool}}
\newcommand{\NatT}{\textit{Nat}}

\newcommand{\Expr}{\ensuremath{\mathit{Expr}}}

\newcommand{\hask}[1]{\emph{#1}\xspace}

\newcommand{\declaration}[0]{\hask{Stream} declaration\xspace}
\newcommand{\declarations}[0]{\hask{Stream} declarations\xspace}
\newcommand{\expression}[0]{\hask{Expression}}
\newcommand{\expressions}[0]{\hask{Expressions}}
\newcommand{\leaf}[0]{\ensuremath{\mathbf{Leaf}}\xspace}

\newcommand{\now}[0]{\ensuremath{\mathbf{Now}}\xspace}
\newcommand{\at}[0]{\ensuremath{\boldsymbol{:\!\!@}}\xspace}
\newcommand{\maxLatency}[0]{\ensuremath{\Varid{maxLatency}}\xspace}
\newcommand{\minBackRef}[0]{\ensuremath{\Varid{minBackRef}}\xspace}
\newcommand{\instant}[0]{\ensuremath{\Conid{Instant}}\xspace}
\newcommand{\instants}[0]{\ensuremath{\Conid{Instant}}s\xspace}
\newcommand{\system}[0]{\hask{Sequence}}
\newcommand{\sequence}[0]{\hask{Sequence}}

\newcommand{\sem}[1]{\llbracket #1\rrbracket}

\newcommand{\hide}[1]{}
\newcommand{\mysize}{\small}

\newcommand{\TRUE}{\textit{True}}
\newcommand{\FALSE}{\textit{False}}

\newcommand{\Thanks}{\thanks{This work was funded in part by Madrid
    Regional Government project ``S2018/TCS-4339 (BLOQUES-CM)'' and by
    Spanish National Project ``BOSCO (PGC2018-102210-B-100)''.}}

\title{Declarative Stream Runtime Verification (hLola)\Thanks}

\author{Mart\'in Ceresa\inst{1}\orcidID{0000-0003-4691-5831} \and Felipe Gorostiaga\inst{1,2,3}\orcidID{0000-0002-3478-3408} \and C\'esar S\'anchez\inst{2}\orcidID{0000-0003-3927-4773}}
\institute{
  CIFASIS, Argentina \and IMDEA Software Institute, Spain  \and Universidad Polit\'ecnica de Madrid, Spain
 }

\begin{document}

\maketitle

\begin{abstract}
  Stream Runtime Verification (SRV) is a formal dynamic analysis
  technique that generalizes runtime verification algorithms from
  temporal logics like LTL to stream monitoring, allowing the
  computation of richer verdicts than Booleans (quantitative values or
  even arbitrary data).
  The core of SRV algorithms is a clean separation between temporal
  dependencies and data computations.
  In spite of this theoretical separation previous engines include
  ad-hoc implementations of just a few data types, requiring complex
  changes in the tools to incorporate new data types.
  
  In this paper we present a solution as a Haskell embedded domain
  specific language that is easily extensible to arbitrary data types.
  The solution is enabled by a technique, which we call \emph{lift
    deep embedding}, that consists in borrowing general Haskell types
  and embedding them transparently into an eDSL.
  This allows for example the use of higher-order functions to
  implement static stream parametrization.
  We describe the Haskell implementation called \hlola and illustrate
  simple extensions implemented using libraries, which would require
  long and error-prone additions in other ad-hoc SRV formalisms.
\end{abstract}

\section{Introduction}
\label{sec:introduction}

In this paper we study the problem of implementing a truly generic
Stream Runtime Verification (SRV) engine, and show a solution using an
embedded domain specific language (eDSL) based on borrowing very
general types from the host language into the SRV language and then
applying a deep embedding.

Runtime Verification
(RV)~\cite{havelund05verify,leucker09brief,bartocci18lectures} is an
area of formal methods for reactive systems that analyses dynamically
one trace of the system at a time.
Compared to static techniques like model
checking~\cite{clarke99model}
RV sacrifices completeness to obtain an applicable and formal
extension of testing and debugging.
Monitors are generated from formal specifications which then inspect a
single trace of execution at a time.
Early RV languages were based on logics like
LTL~\cite{manna95temporal} or past LTL adapted for finite
paths~\cite{bauer11runtime,eisner03reasoning,havelund02synthesizing}.
Other approaches followed, based on regular
expressions~\cite{sen03generating}, rule based
languages~\cite{barringer04rule}, or rewriting~\cite{rosu05rewriting}.
These specification languages come from static verification, where
decidability is key to obtain algorithmic solutions to decision
problems like model checking.
Therefore, the observations and verdicts are typically Boolean values.
Stream Runtime Verification~\cite{dangelo05lola,sanchez18online}
starts from the observation that most monitoring algorithms for logics
from static verification can be generalized to richer observations and
outcomes (verdicts), by generalizing the datatypes of the individual
internal operations of the monitors.
Languages for SRV, pioneered by \lola~\cite{dangelo05lola}, describe
monitors declaratively via equations that relate streams of input and
streams of output, offering a clean separation between the time
dependencies and the concrete operations.
The temporal part is a sequence of operations on abstract data,
mimicking the steps of the algorithms for temporal logics.
Each individual operation can then be performed on a datatype
implementation, obtaining monitors for arbitrary data.
Offset expressions allow us to refer to stream values in different
moments of time, including future instants (that is, SRV monitors
need not be causal).

Most previous SRV
developments~\cite{convent18tessla,leucker18tessla,gorostiaga18striver,faymonville19streamlab}
focus on efficiently implementing the temporal engine, promising that
the clean separation between time and data allows incorporating
off-the-shelf datatypes easily.
However, in practice, adding a new datatype requires modifying the
parser, the internal representation, and the runtime system that keeps
track of offset expressions and partially evaluated expressions.
Consequently, these tools only support a limited hard-wired collection
of datatypes.
In this paper, we give a general solution to this problem via a
Haskell eDSL, resulting in the language \hlola\footnote{available open
  source at \url{http://github.com/imdea-software/hlola}}, whose
engine implements a generic SRV monitoring algorithm that works for
general datatypes.

Typically, a DSL is designed as a complete language, first defining
the types and terms of the language (this is, the underlying theory),
which is then im\-ple\-men\-ted---either as an eDSL or as a standalone
DSL---, potentially mapping the types of the DSL into types of the
host.
However, our intention with \hlola is to have a language where
datatypes are not decided upfront but can be added on demand without
requiring any re-implementation.
For this reason, \hlola borrows (almost) arbitrary types from the host
system and then embeds all these borrowed types, so \hlola is agnostic
from the stream types (even types added in the future).
Even though this technique has been somewhat part of the folklore of
modern Haskell based eDSLs
(e.g.~\cite{westphal20implementing}), this is a
novel approach to build runtime verification engines.
We called this technique a \emph{lift deep embedding}, which consists
of (1) lifting the types and values of the host language into the
generic DSL using generic programming, and (2) deep embedding the
resulting concrete DSL into the host language.
This technique allows us to incorporate Haskell datatypes into \hlola,
and enables the use of many features from the host language in the
DSL.
For example, we use higher-order functions to describe transformations
that produce stream declarations from stream declarations, obtaining
static parameterization for free.
In turn, libraries collect these transformers, which allows defining
in a few lines new logics like LTL, MTL, etc. or quantitative
semantics for these logics.
Haskell type-classes allow us to implement \emph{simplifiers} which can
compute the value of an expression without resolving all its
sub-expressions first.
If the unevaluated sub-expressions contain future offset references,
the engine may anticipate verdicts ahead of time.
Implementing many of these in previous SRV systems has required to
re-invent and implement features manually (like macro expansions or
ad-hoc parameterization).
We use polymorphism both for genericity (to simplify the engine
construction) and to enable the description of generic stream
specifications, which, again, is not allowed by previous SRV engines.
Finally, we also exploit features present in Haskell to offer IO for
many stream datatypes for free.

\paragraph{Related work.}
SRV was pioneered by \lola~\cite{dangelo05lola} for monitoring
synchronous systems only supporting Integers and Booleans.
Copilot~\cite{pike10copilot} is a Haskell implementation that offers a
collection of building blocks to transform streams into other streams,
but Copilot does not offer explicit time accesses (and in particular
future accesses).
\textsc{Lola} 2.0~\cite{faymonville19streamlab} extends \lola with
special constructs for runtime parameterization and real-time
features.
TeSSLa~\cite{convent18tessla} and Striver~\cite{gorostiaga18striver}
are two modern SRV approaches that target real-time event streams.
All these languages still support only limited hard-wired datatypes.

RV and SRV overlap with other areas of research.
Synchronous languages --like Esterel~\cite{berry00foundations} or
Lustre~\cite{halbwachs87lustre}-- are based on data-flow.
These languages force causality because their intention is to describe
systems and not observations or monitors, while SRV removes the
causality assumption allowing the reference to future values.
In Functional Reactive Programming (FRP)~\cite{eliot97functional}
reactive behaviors are defined using the building blocks of functional
programming.
An FRP program describes a step of computation, a reaction when new
input values (events) are available, thus providing implicitly the
dependency of the output streams at the current point from input
streams values.
Again, the main difference is that FRP programs do not allow explicit
time references and that the dependencies are causal (after all, FRP
is a programming paradigm).
In comparison, FRP allows immediately all the features of the
programming language without needing the solution proposed in this
paper.
It would be interesting to study the opposite direction that we solve
in this paper: how to equip FRP with explicit times and non-causal
future references.
Also, FRP does not typically target resource calculation, while this
is a main concern in RV (and in SRV).

\paragraph{Contributions.}
In summary, the contributions of the paper are:
 \begin{inparaenum}[(1)]
\item An implementation of SRV, called \hlola, based on an eDSL that
  exploits advanced features of Haskell to build a generic engine.
  A main novelty of \hlola as an SRV implementation is the use of a
  lift deep embedding to gain very general types without costly
  implementations.
  Section~\ref{sec:implementation} describes the runtime system of
  \hlola.
\item An implementation of many existing RV specification languages
  (including LTL, MT-LTL and MTL) in \hlola, which illustrates the
  simplicity of extending the language.
  This is shown in Section~\ref{sec:libraries}.
\item A brief empirical evaluation, which suggests that the \hlola
  engine executes using only the theoretically predicted resources,
  shown in Section~\ref{sec:empirical}.
   \end{inparaenum}
%%% %include related.tex
\section{Preliminaries}
\label{sec:preliminaries}

We briefly introduce SRV using \lola (see~\cite{sanchez18online}) and
then present the features of Haskell as a host language that we use to
implement \hlola.

\subsection{Stream Runtime Verification: Lola}
Intuitively speaking, \lola is a specification language and a
monitoring algorithm for synchronous systems.
\lola programs describe monitors by expressing, in a declarative
manner, the relation between output streams and input streams.
Streams are finite sequences of values, for example, a Boolean stream
is a sequence of Boolean values.
The main idea of SRV is to cleanly separate the temporal dependencies
from the data computation.

For the data, monitors are described declaratively by providing one
expression for each output stream.
Expressions are terms from a multi-sorted first order theory, given by
a first-order signature and a first-order structure.
A theory is a finite collection of interpreted \emph{sorts} and a
finite collection of interpreted function symbols.
Sorts are interpreted in the sense that each sort is associated with a
\emph{domain}, for example the domain of sort $\BoolT$ is the set of
values $\{\true,\false\}$.
For the purpose of this paper we use sorts and types interchangeably,
as we use Haskell types to implement \lola sorts.
Function symbols are interpreted, meaning that $f$ is both (1) a
constructor to build terms; and (2) a total function (the
interpretation) used to evaluate and obtain values of the domain of
the return sort.
For example, natural numbers uses two sorts (\NatT{} and \BoolT),
constant function symbols $0$, $1$, $2$, $\ldots$ of sort \NatT{}, and
$\textit{True}$ and $\textit{False}$ of type $\BoolT$, as well as
functions $+$, $*$, $\cdots$ $\NatT\times \NatT \Into \NatT$ and
predicates $<$, $\leq$, \ldots, that are symbols that return $\BoolT$.
We assume that our theories include equality, and also that for every
sort $T$ there is a ternary function
$\IfThenElse{\cdot}{\cdot}{\cdot}$ that returns a value of sort $T$ given
a Boolean and two arguments of sort $T$.
We use $e:T$ to represent that $e$ has sort $T$.

Given a set $Z$ of (typed) \emph{stream variables}, \emph{offset
  expressions} are $v[k,d]$ where $v$ is a stream variable, $d:T$ is a
constant and $k$ is an integer number.
For example, $x[-1,\false]$ is an $\BoolT$ offset expression
and $y[+3,5]$ is a $\NatT$ offset expression.
The intended meaning of $v[k,d]$ is to represent, at time $n$, the
value of the stream $v$ at time $n+k$.
The second argument $d$ indicates the default value to be used beyond
the time limits.
When it is clear from the context, we use $v$ to refer to the offset
expression $v[0]$ (that does not need a default value).
The set of \emph{stream expressions} over a set of variables $Z$
(denoted $\Expr(Z)$) is the smallest set containing $Z$ and all
offset expressions of variables of type $Z$, that is closed under
constructor symbols of the theory used.
For example $(x[-1,\false] \Or x)$ and $(y + y[+3,5] * 7)$ are stream
expressions.

A \emph{\lola specification} consists of a set $\{s_1,s_2\ldots\}$ of
input stream variables and a set $\{t_1,t_2\ldots\}$ of output stream
variables, and one \emph{defining expression} $t_i=\textit{exp}_i$ per
output variable over the set of input and output streams, including
$t_i$ itself.

\begin{example}
  \label{langdesign:once_s_spec}
  \label{ex:once_s_spec}
  The following is a \lola specification with input stream variable
  $s:\BoolT$ and output stream variable $\textit{once\_s}:\BoolT$:

{\mysize  \begin{lstlisting}[language=LOLA]
    input   bool s
    output bool once_s = once_s [-1,false] || s
  \end{lstlisting}}

\noindent This example corresponds to the LTL formula $\Once s$.
The following specification counts how many times $s$ was
$\textit{True}$ in the past (\verb+toint+ is the function that returns
$0$ for $\textit{False}$ and $1$ for $\textit{True}$):
  {\mysize \begin{lstlisting}[language=LOLA] input
    output int n_once_s = n_once_s [-1,0] + toint(s)
\end{lstlisting}}
\end{example}

A valuation of a specification associates a stream of length $N$ to
each of its stream variables, all of which are of the same length.
Given a stream $\sigma_i$ for each input stream variable $s_i$ and a
stream $\tau_i$ for each output stream variable $t_i$ in a
specification, every expression $e$ can be assigned a stream $\sem{e}$
of length $N$.
For every $j = 0 \ldots N-1$:
\begin{compactitem}
\item $\sem{c}(j)=c$ for constants;
\item $\sem{s_i}(j)=\sigma_i(j)$ and $\sem{t_i}(j)=\tau_i(j)$ for stream variables;
\item
  $\sem{f(e_1,\ldots,e_n)}(j)=f(\sem{e_1}(j),\ldots,\sem{e_n}(j))$; and
 \item $\sem{v[k,d]}(j)=\sem{v}(j+k)$ if $0\leq j+k<N$, and
   $\sem{v[k,d]}(j)=d$ otherwise.
\end{compactitem}
We say that a valuation is an evaluation model, if
$\sem{t_i}=\sem{e_i}$ for each output variable $t_i$, that is, if
every output stream satisfies its defining equation.
The dependency graph is the graph of offset dependencies between
variables, and can be used to rule out cycles in specifications to
guarantee that every specification has a unique output for every
input.

One very important aspect of SRV is its ability to analyze
specifications and automatically calculate the necessary resources.
A monitor is \emph{trace-length independent} if it operates with an
amount of memory (and of processing time per input event) that does
not depend on the length of the trace.
Many logics admit trace-length independent algorithms for their past
fragments, like for example LTL and TLTL~\cite{bauer11runtime} and
MTL~\cite{thati05monitoring}.
%\cite{thati05monitoring,ho14online,basin17almost}.
%
The notion of \emph{efficient monitorability} in
SRV~\cite{dangelo05lola,sanchez18online}, defined as the absence of
positive (future) cycles in the dependency graph, guarantees a
trace-length independent monitor.
The dependency graph can also be used to build efficient runtime
systems, by determining when a value stream variable is guaranteed to
be resolved (the latency) and when a value can be removed because it
will no longer be necessary (the back-reference).
See~\cite{sanchez18online} for longer formal definitions.
 % we need these be separate files
\label{subsec:lola}

\subsection{Haskell as a host language for an eDSL}
\label{sec:intro:haskelledsl}

An \emph{embedded Domain Specific Language}\cite{Hudak:EDSL} (eDSL) is
a DSL that can be used as a language by itself, and also as a library
in its host programming language.
An eDSL inherits the language constructs of its host language and adds
domain-specific primitives.
In this work we implemented \hlola as an eDSL in Haskell.
In particular, we use Haskell's features as host language to implement
static parameterization (see Section~\ref{sec:staticparam}), a
technique that allows the programmatic definition of specifications.
This is used to extend \hlola to support many temporal logics proposed
in RV.
Other SRV implementations, in their attempt to offer expressive data
theories in a standalone tool, require a long and costly
implementation of features that are readily available in higher-order
expressive languages like Haskell.
Using an eDSL, we can effectively focus our development efforts on the
temporal aspects of \lola.

We describe in the next section the \emph{lift deep embedding}, which
allows us to lift Haskell datatypes to \lola and then to perform a
single deep embedding for all lifted datatypes.
This technique fulfills the promise of a clean separation of time and
data and eases the extensibility to new data theories, while keeping
the amount of code at a minimum.
Additionally, using eDSLs brings benefits beyond data theories,
including leveraging Haskell's parsing, compiling, type-checking, and
modularity.
The drawback is that specifications have to be compiled with a Haskell
compiler, but once a specification is compiled, the resulting binary
is agnostic of the fact that an eDSL was used.
Therefore, any target platform supported by Haskell can be used as a
target of \hlola.
Moreover, improvements in the Haskell compiler and runtime systems
will be enjoyed seamlessly, and new features will be ready to be used
in \hlola.
%

%%%%%%%%%%%%%%%%%%%%%%%%%%%%%%%%%%%%%%%%%%%%%%%%%%%%%%%%%%%%%%%%%%%%%%%%%%%%%%%%
%%%%%%%%%%%%%%%%%%%% Haskell
%%%%%%%%%%%%%%%%%%%%%%%%%%%%%%%%%%%%%%%%%%%%%%%%%%%%%%%%%%%%%%%%%%%%%%%%%%%%%%%%

Haskell~\cite{Haskell2010} is a pure statically typed functional
programming language that has been reported to be an excellent host
language for eDSLs~\cite{GillHaskEDSL}.
Functions are values, and function application is written simply as a
blank space without parentheses, which helps eDSLs look cleaner.
Haskell also allows custom parametric polymorphic datatypes, which
% \begin{itemize}
% \item
  eases the definition of new data theories, and
% \item
  enables us to abstract away the types of the streams,
% \end{itemize}
effectively allowing the expression of generic specifications.

Haskell's ecosystem provides a plethora of frameworks for generic
programming~\cite{HaskellGenericProgramming}.
In particular, our engine implementation uses the \ensuremath{\Conid{Typeable}} class to
incorporate new types without modification.
However, we do not perform any kind of traversal over generic data, we
employ the \ensuremath{\Conid{Typeable}} class as a mechanism to hide concrete types and
implement heterogeneous lists.
Members of the \ensuremath{\Conid{Typeable}} class have an associated type
representation, which can be compared, and therefore employed to
define a \ensuremath{\Conid{Dynamic}} datatype (which hides a \ensuremath{\Conid{Typeable}} datatype), and
to define a type-safe cast operation.
New datatypes developed by the active Haskell community can be
incorporated immediately into \hlola.
The datatype members of the \ensuremath{\Conid{Typeable}} class encompass all sorts that
are used in practice in SRV.

Haskell is declarative and statically typed, just like \lola.
In \lola, functions are functions in the mathematical sense, that is,
they do not have side effects.
\lola does not make assumptions about when these functions will be
called, and guarantees that a function yields the same result when
applied to the same arguments twice.
This is aligned with the Haskell purity of (total) functions.

Another feature that improves syntax readability is Haskell type
classes, which allows overloading methods.
We can redefine functions that are typically native in other
languages, such as Boolean operators \ensuremath{(\mathrel{\vee})} and \ensuremath{(\mathrel{\wedge})}, and the
arithmetic operators \ensuremath{(\mathbin{+})}, \ensuremath{(\mathbin{-})} and \ensuremath{(\mathbin{*})}, as well as define and use
custom infix operators.
Such definitions are possible by extensions made by the de-facto
Haskell compiler, GHC~\cite{GHC}.
Haskell has let-bindings, list comprehensions, anonymous functions,
higher-order, and partial function application, all of which improves
specification legibility.
Finally, \hlola uses Haskell's module system to allow modular
specifications and to build language extensions.

\section{Implementation}
\label{sec:implementation}
%\vspace{-1em}
\subsection{Language design}
\label{sec:langdesign}

We model input and output stream variables using:
\begin{compactitem}
\item \hask{Input} \declarations, which model \lola's input variables
  simply as a \emph{name}. During evaluation, the engine can look up a
  name in the environment and fetch the corresponding value at a
  required time instant.
\item \hask{Output} \declarations, which model output streams in
  \lola.
  These declarations bind the name of the stream with its \expression,
  which represents the defining expression of a \lola output stream.
\end{compactitem}
Revisiting the \lola specification in Ex.~\ref{ex:once_s_spec}, in
\hlola, \ensuremath{\Varid{s}} will be an \hask{Input} \declaration and \ensuremath{\Varid{once\char95 s}} an
\hask{Output} \declaration.

We seek to represent many theories of interest for RV and to
incorporate new ones transparently, so we abstract away concrete types
in the eDSL.
For example, we want to use the theory of \emph{Boolean} without
adding the constructors that a usual deep embedding would require.
To accomplish this goal we revisit the very essence of functional
programming.
Every expression in a functional language---as well as in
mathematics---is built from two basic constructions: \emph{values} and
\emph{function applications}.
Therefore, to implement our SRV engine we use these two constructions,
plus two additional stream access primitives to capture offset
expressions.
The resulting datatype is essentially a
de-functionalization~\cite{Defuntionalization} of the applicative
interface.
There is a limitation that some Haskell datatypes cannot be handled
due to the use of \ensuremath{\Conid{Dynamic}} and \ensuremath{\Conid{Typeable}}, which we introduce within
the engine to get a simple way to implement generic programming while
preserving enough structure.
However, this is not a practical limitation to represent theories and
sorts of interests for monitoring.
We define expressions in Haskell as a parametric datatype
\ensuremath{\Conid{Expr}} with a polymorphic argument \ensuremath{\Varid{domain}}.
An \ensuremath{\Varid{e}\mathbin{::}\Conid{Expr}\;\Varid{domain}} represents an expression \ensuremath{\Varid{e}} over the domain
\ensuremath{\Varid{domain}}.
The generic \ensuremath{\Varid{domain}} is automatically instantiated at static time by
the Haskell compiler, effectively performing the desired lifting of
Haskell datatypes to types of the theory in \hlola.
For example, the use of \ensuremath{\Conid{Expr}\;\Conid{Int}} will make the compiler instantiate
\ensuremath{\Varid{domain}} as \ensuremath{\Conid{Int}}.
The resulting concrete \expressions constitute a typical deeply
embedded DSL.
We call this two step technique a \emph{lift deep embedding}.
This technique avoids defining a constructor for all elements in the
data theory, making the incorporation of new types transparent.

Here we present in more detail the \ensuremath{\Conid{Expr}} construction in
Haskell.
The first two constructors (\ensuremath{\mathbf{Leaf}} and \ensuremath{\mathbf{App}}) are the \emph{data
constructions} of the language, which are aligned with the notions of
de-functionalization mentioned above, and allow encoding terms from a
\lola theory seamlessly.
The other two constructors (\now and (\at)) represent the offset
expressions:
\begin{compactitem}
\item The constructor \ensuremath{\mathbf{Leaf}\mathbin{::}\Conid{Typeable}\;\Varid{a}\Rightarrow \Varid{a}\to \Conid{Expr}\;\Varid{a}} models an
  element of the theory.
\item
  The constructor \ensuremath{\mathbf{App}\mathbin{::}(\Conid{Typeable}\;\Varid{a},\Conid{Typeable}\;\Varid{b},\Conid{Typeable}\;(\Varid{b}\to \Varid{a}))\Rightarrow } \\ \ensuremath{\Conid{Expr}\;(\Varid{b}\to \Varid{a})\to \Conid{Expr}\;\Varid{b}\to \Conid{Expr}\;\Varid{a}} represents the application of
  a \emph{function} \expression to a \emph{value} \expression.
\item
A term \ensuremath{\mathbf{Now}\mathbin{::}\Conid{Stream}\;\Varid{a}\to \Conid{Expr}\;\Varid{a}}  represents the value of a stream in the current
  instant.
\item The \emph{at} infix constructor, \ensuremath{(\boldsymbol{:\!\!@})\mathbin{::}\Conid{Stream}\;\Varid{a}\to (\Conid{Int},\Conid{Expr}\;\Varid{a})\to \Conid{Expr}\;\Varid{a}} models future and past offset expressions, specifying the
  default value to use if the access falls off the trace
\end{compactitem}
These constructions allow us to lift operations from domain values to
\expressions directly.
For example, we can create an \expression that represents the sum of two
\hask{Expr Int} without defining a dedicated type of \expression.

Similarly, we define the \declarations in Haskell as a parametric
datatype \ensuremath{\Conid{Stream}} with a polymorphic argument \ensuremath{\Varid{domain}}.
\begin{compactitem}
\item The \ensuremath{\mathbf{Input}\mathbin{::}(\Conid{FromJSON}\;\Varid{a},\Conid{Read}\;\Varid{a},\Conid{Typeable}\;\Varid{a})\Rightarrow \Conid{String}\to \Conid{Stream}\;\Varid{a}} constructor represents an input stream, and
associates the name of the stream to the type of its values.
\item
The \ensuremath{\mathbf{Output}\mathbin{::}\Conid{Typeable}\;\Varid{a}\Rightarrow (\Conid{String},\Conid{Expr}\;\Varid{a})\to \Conid{Stream}\;\Varid{a}} constructor
represents an output stream, and associates the name of the stream to
the type of its values and its defining \expression, of the same type.
\end{compactitem}

\noindent The \lola specification from Ex.~\ref{langdesign:once_s_spec} can be
written in \hlola as follows:
{\mysize
  \begin{hscode}\SaveRestoreHook
\column{B}{@{}>{\hspre}l<{\hspost}@{}}%
\column{3}{@{}>{\hspre}l<{\hspost}@{}}%
\column{E}{@{}>{\hspre}l<{\hspost}@{}}%
\>[B]{}\Varid{once\char95 s}\mathbin{::}\Conid{Stream}\;\Conid{Bool}{}\<[E]%
\\
\>[B]{}\Varid{once\char95 s}\mathrel{=}\mathbf{Output}\;(\text{\ttfamily \char34 once\char95 s\char34},\mathbf{App}\;(\mathbf{App}\;(\mathbf{Leaf}\;(\mathrel{\vee}))\;\Varid{prevOnce\char95 s})\;(\mathbf{Now}\;\Varid{s})){}\<[E]%
\\
\>[B]{}\hsindent{3}{}\<[3]%
\>[3]{}\mathbf{where}\;\Varid{s}\mathrel{=}\mathbf{Input}\;\text{\ttfamily \char34 s\char34}{}\<[E]%
\\
\>[B]{}\hsindent{3}{}\<[3]%
\>[3]{}\qquad\quad\;\Varid{prevOnce\char95 s}\mathrel{=}\Varid{once\char95 s}\boldsymbol{:\!\!@}(\mathbin{-}\mathrm{1},\Conid{False}){}\<[E]%
\ColumnHook
\end{hscode}\resethooks
}
\noindent The expression of \ensuremath{\Varid{once\char95 s}} takes the application of the
(data theory) function \ensuremath{(\mathrel{\vee})} to the value of \ensuremath{\Varid{once\char95 s}} at $-1$,
using \ensuremath{\Conid{False}} as the default value, and applying the result to the
current value of \ensuremath{\Varid{s}}.
We define the infix operator \ensuremath{(\boldsymbol{=:})} that builds an output stream from
a name and an expression, and override the Boolean operator \ensuremath{\vee};
and the \hlola \hask{Output} \declaration looks almost like a \lola
expression:\\
{\mysize
\indent \ensuremath{\qquad\Varid{once\char95 s}\mathrel{=}\text{\ttfamily \char34 once\char95 s\char34}\boldsymbol{=:}\Varid{once\char95 s}\boldsymbol{:\!\!@}(\mathbin{-}\mathrm{1},\Conid{False})\mathrel{\vee}\mathbf{Now}\;\Varid{s}}
}
%
% \vspace{-2em}
\subsection{Static analysis}
\label{subsec:static}
Not every grammatically correct \lola specification is valid.
Some errors like syntactic errors, missing references and type
mismatches can be checked by the Haskell compiler.
But to guarantee that a specification is well-defined we also need
to examine the dependency graph to check that it does not contain
closed paths of weight zero.
This will ensure that the value of a stream at any point does not
depend on itself.
We first convert every \expression \ensuremath{\Varid{a}} and \ensuremath{\Conid{Stream}\;\Varid{a}} to
their equivalent \expression and \declaration of \ensuremath{\Conid{Dynamic}}, so
\declarations and \expressions of different types can be mixed and
used in the same specification.
Then we obtain the dependency graph by traversing the stream
definitions in the specification recursively.
One drawback of this approach is that the Haskell type-system can no
longer track the original type of an expression, but this step is made
after Haskell has type-checked the specification, guaranteeing that
the engine is forgetting the type information of a well-typed
specification.
The engine keeps the information on how to parse the input streams and
how to show output values given a stream name, safely casting from and
into \ensuremath{\Conid{Dynamic}}, and avoiding type mismatches when converting from
dynamically-typed objects.
We make the following claim:
\begin{claim}
  Every conversion from a Dynamic \expression within the \hlola engine
  returns a value \expression of the expected type.
\end{claim}

The proof of this claim can be done using Liquid
Haskell~\cite{vazou14liquid} and is ongoing research beyond the scope
of this paper.
Assuming the claim above, a runtime type error can only be produced
when processing an input event whose value is not of the expected
type.

During this stage, the tool also calculates the minimum weight of
the paths in the dependency graph, a non-positive value that we call
\emph{minimum back reference} and note \minBackRef, along with the
maximum weight of the edges, which we call \emph{latency} and note
\maxLatency.
The dependency graph of the
\begin{wrapfigure}[8]{c}{0.23\textwidth}
  \vspace{-2.3em}
  \centering
    \includegraphics[scale=0.8]{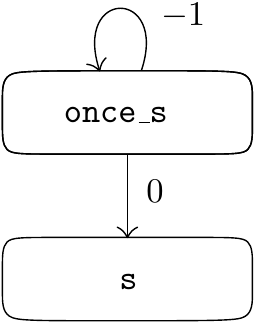}
\end{wrapfigure}
specification in Ex.~\ref{langdesign:once_s_spec} is
shown on the right.
The \minBackRef is $-1$, because \ensuremath{\Varid{once\char95 s}} depends on the previous
value of itself, and the \maxLatency is $0$ because there are no
references to future values of streams.  The values of \minBackRef and
\maxLatency indicate that the engine will only keep the values of the
streams at the present and previous instants.
\subsection{Runtime System}
\label{subsec:runtime}
We now describe some key internal datatypes used in the implementation
of the execution engine.
An \instant is a map that binds the name of a stream to an \expression.
Given a specification with $m$ streams $s_1,\dots,s_m$, an \instant can
be interpreted as a vector of size $m$.
A \system is an ordered collection of \instants, one of which is said to be ``in
focus''.
The \instants in the past of the one in focus are stored in the
\system in an array of size $(\maxLatency - \minBackRef)$, which
limits the amount of memory that the engine can consume.
On the other hand, the \instants in the future of the one in focus are stored as
a list.
Even though this list can be (implicitly) as long as the full trace,
the elements in the list will not actually exist in memory until they
are needed to compute a value, due to the laziness of Haskell
evaluation.
We can think of a \sequence as a matrix of expressions, where each
column is an \instant vector, and one of them is in focus.
The evaluation of a specification with $m$ streams over $n$ instants
is conceptually an $n\times m$ matrix.

Given a specification and a list of values, we first create a \sequence with an
empty past and the focus on the first instant.
In this \sequence, the value of the cell $(s_i,n)$ in the \system matrix for an
\emph{input} stream $s_i$ and instant $n$, is a \leaf containing the value read
for the stream of $s_i$ at time instant $n$.
Similarly, the value of every \emph{output} stream $t_j$ and instant $n$ is
the defining \expression for $t_j$ in the specification, waiting to be
specialized and evaluated.
Note that these values do not actually exist in memory until they are
needed.
The goal of the engine is to compute a \leaf expression (this is, a
ground value) at every position in the matrix, particularly for output
streams.

Starting from the initial state, the engine will solve every output stream at
the instant in focus, and then move the focus one step forward.
This algorithm guarantees that all elements in the past of the focus are leaves.
\begin{wrapfigure}[8]{c}{0.57\textwidth}
 \vspace{-3em}
  \centering
  {\small\[
      \begin{blockarray}{cccccc}
        &1 & 2 & 3 & \dots & n \\
        \begin{block}{c(ccccc)}
          s_1~~ & \leaf_{1,1} & \leaf_{1,2} & \leaf_{1,3} & \dots  & \leaf_{1,n} \\
          \vdots & \vdots  & \vdots & \vdots & \ddots & \vdots \\
          s_k~~ & \leaf_{k,1} & \leaf_{k,2} & \leaf_{k,3} & \dots  & \leaf_{k,n} \\
          t_{1}~~ & \leaf_{k+1,1} & \leaf_{k+1,2} & e_{1,3} & \dots  & e_{1,n} \\
          \vdots & \vdots  & \vdots & \vdots & \ddots & \vdots \\
          t_m~~ & \leaf_{k+m,1} & \leaf_{k+m,2} & e_{m,3} & \dots  & e_{m,n} \\
        \end{block}
        \begin{block}{cccccc}
          & & &\bigtriangleup & & \\
        \end{block}
      \end{blockarray}
    \]}
\end{wrapfigure}
The figure on the right illustrates the \system of an 
execution at
time instant $3$, where some of the output expressions
$e_{1,3} \ldots e_{m,n}$ can be leaves too.
At the end of the execution, the focus will be on the last column of
the matrix, and all the elements in the matrix will be leaves.

\newcommand{\ExSequence}{
{\small \[
\begin{blockarray}{ccccc}
&1 & 2 & 3 & \dots \\
\begin{block}{c(cccc)}
s~~ & \leaf~\FALSE & \leaf~\TRUE & \leaf~\FALSE & \dots \\
once\_s~~ & \leaf~\FALSE & \leaf~\TRUE & \leaf~\TRUE & \dots \\
\end{block}
\begin{block}{ccccc}
& & & &\bigtriangleup \\
\end{block}
\end{blockarray}
\]}
}
\begin{example}
  Consider the specification from Ex.~\ref{langdesign:once_s_spec}
  and suppose that the first three elements read for input stream \ensuremath{\Varid{s}}
  are \ensuremath{\Conid{False}}, \ensuremath{\Conid{True}} and \ensuremath{\Conid{False}}.
%
% \[
% \begin{blockarray}{ccccc}
% &1 & 2 & 3 & \dots \\
% \begin{block}{c(cccc)}
% s~~ & False & True & False & \dots \\
% \end{blockarray}
% \]
%
Then, the initial state of our system will be a \sequence containing
the values read as leaves for \ensuremath{\Varid{s}} at the first three instants and the
defining expression (\ensuremath{\Varid{once\char95 s}\boldsymbol{:\!\!@}(\mathbin{-}\mathrm{1},\Conid{False})\mathrel{\vee}\mathbf{Now}\;\Varid{s}}) at every
position in the row for \ensuremath{\Varid{once\char95 s}}:

{\small\[
\begin{blockarray}{ccccc}
&1 & 2 & 3 & \dots \\
\begin{block}{c(cccc)}
s~~ & \leaf~False & \leaf~True & \leaf~False & \dots \\
once\_s~~ & e & e & e & \dots \\
\end{block}
\begin{block}{ccccc}
&\bigtriangleup & & & \\
\end{block}
\end{blockarray}
\]}

\noindent where $e = $ \ensuremath{\Varid{once\char95 s}\boldsymbol{:\!\!@}(\mathbin{-}\mathrm{1},\Conid{False})\mathrel{\vee}\mathbf{Now}\;\Varid{s}}.
After the first iteration, the first instant of \ensuremath{\Varid{once\char95 s}} is resolved.
Since \ensuremath{\Varid{once\char95 s}} does not exist at instant $-1$, the default value
\ensuremath{\Conid{False}} is used to compute the disjunction with the current value of
\ensuremath{\Varid{s}}, resulting in \ensuremath{\Conid{False}}.
As a consequence, the expression of \ensuremath{\Varid{once\char95 s}} at position $1$ is replaced with
$\leaf~$\ensuremath{\Conid{False}}.
The state becomes:

{\small\[
\begin{blockarray}{ccccc}
&1 & 2 & 3 & \dots \\
\begin{block}{c(cccc)}
s~~ & \leaf~False & \leaf~True & \leaf~False & \dots \\
once\_s~~ & \leaf~False & e & e & \dots \\
\end{block}
\begin{block}{ccccc}
& &\bigtriangleup & & \\
\end{block}
\end{blockarray}
\]}

In the next iteration, with the second instant in focus, the
defining expression of \ensuremath{\Varid{once\char95 s}} is evaluated using the value
of \ensuremath{\Varid{once\char95 s}} at instant $1$, which is \ensuremath{\Conid{False}}, and operated via
disjunction with the value of \ensuremath{\Varid{s}} at instant $2$, which is \ensuremath{\Conid{True}}, and
therefore the evaluation becomes \ensuremath{\Conid{True}}.
Since \ensuremath{\Conid{True}} is the absorbing element of the disjunction, the value of
\ensuremath{\Varid{once\char95 s}} will be \ensuremath{\Conid{True}} from instant $2$ onwards, as shown below.
% on the right.

\begin{figure}[htp]
 \vspace{-1.5em}
  \ExSequence
  \vspace{-2em}
\end{figure}
% \begin{wrapfigure}[4]{c}{0.58\textwidth}
% \ExSequence
% \end{wrapfigure}
%
\vspace{-2em} \qed
\end{example}

The output streams will be calculated and output incrementally while
new data is retrieved for the input streams.
The engine will block when it needs the value of an input stream that has not
been provided yet.
These characteristics of the Haskell runtime system allow the monitor to run
online processing events from the system under analysis on the fly, or offline
over dumped traces.
A language that offers means to define new datatypes must not only
provide the constructs to define them, but it also must implement the
encoding and decoding of custom datatypes.
Extensible encoding and decoding of datatypes in the theory is not
trivial and might account for a large portion of the codebase.
As an eDSL, \hlola can rely upon Haskell's facilities to define how to
encode and decode \ensuremath{\Conid{Typeable}} datatypes, sometimes even automatically
from their definitions.
This class encompasses many of the datatypes that are used in
practice to encode values (observations and verdicts) when monitoring
systems.

Input events are fed to \hlola in JSON format, where each line is a
string representation of a JSON object with one field per input
stream.
The types of the input streams have to be instances of the \ensuremath{\Conid{FromJSON}}
class, meaning that a value of the corresponding type can be
constructed from a serialized JSON \ensuremath{\Conid{Object}}.
Output streams must be instances of the \ensuremath{\Conid{ToJSON}} class, which means
that we can get a JSON \ensuremath{\Conid{Object}} from a value of the corresponding
type.

Haskell allows defining custom datatypes via the \ensuremath{\mathbf{data}} statement.
Once defined, these types can be used just like any other type in Haskell.
Most of the times, we can use Haskell's \ensuremath{\mathbf{deriving}} mechanism to make
our custom types instances of the corresponding classes, if needed.
Section~\ref{sec:libraries} contains examples of custom datatypes for
input values.
%
%\vspace{-1em}
% %include inout.tex
%
% \vspace{-1em}
\subsection{Additional Features}
\label{sec:implementation:features}

The use of Haskell as a host language eases the implementation of many
useful features of SRV in \hlola.
We show here two examples: anticipation and parameterized streams.

\subsubsection{Anticipation}
\label{sec:anticipation}
Input event streams represent the trace of observations of a system, and
output streams encode a property to be evaluated dynamically.
The principle of \textit{anticipation}, as presented
in~\cite{dong08impartial}, states that once every (infinite)
continuation of a finite trace leads to the same verdict, then the
finite trace can safely evaluate to this verdict.
This principle can be trivially implemented when functions know all
their arguments, but it is not always possible to anticipate what the output of
the function will be when some of the arguments will only be known in the
future.
Nevertheless, there are cases where a function can be evaluated with
just a subset of its arguments.
This property of some functions can be used to compute their values as
soon as all the relevant information is retrieved, avoiding waiting
for input values that are not strictly necessary to evaluate the
function.
This idea effectively brings us closer to strict anticipation as
defined above.

The circumstances under which a function can be computed with missing
arguments is data-specific information.
Typical SRV implementations provide simplifications for some functions
in the covered theories, but do not offer a way to provide new
simplifications to their theories.
Instead, we provide a framework to keep the simplifications
extensible.
To allow the use of functions off-the-shelf as well as simplifiable
functions, we define a new datatype and a class of which the Haskell
function constructor \ensuremath{(\to )} is an instance, shown below:

\vspace{1em}

{\mysize 
% \begin{code}
 \ensuremath{\mathbf{data}\;\Conid{LFunction}\;\Varid{a}\;\Varid{b}\mathrel{=}\Conid{Pure}\;(\Varid{a}\to \Varid{b})\mid \Conid{Simplifier}\;(\Conid{Maybe}\;\Varid{a}\to \Conid{Maybe}\;\Varid{b})}

\ensuremath{\mathbf{class}\;\Conid{ILFunction}\;\Varid{x}\;\mathbf{where}\;\Varid{toLFunction}\mathbin{::}\Varid{x}\;\Varid{a}\;\Varid{b}\to \Conid{LFunction}\;\Varid{a}\;\Varid{b}}

\ensuremath{\mathbf{instance}\;\Conid{ILFunction}\;(\to )\;\mathbf{where}\;\Varid{toLFunction}\mathrel{=}\Conid{Pure}}
% \end{code}
}

\vspace{1em}

We then generalize the type of the function application constructor
\ensuremath{\mathbf{App}\mathbin{::}\Conid{Expression}\;(\Varid{f}\;\Varid{b}\;\Varid{a})\to \Conid{Expression}\;\Varid{b}\to \Conid{Expression}\;\Varid{a}}, under the
constraint that \ensuremath{\Varid{f}} be a member of the class \ensuremath{\Conid{ILFunction}}.
In this way, users of the eDSL can define their own simplifiable
functions using the \ensuremath{\Conid{Simplifier}} constructor, or just use off-the-shelf
functions seamlessly; which will automatically be applied the \ensuremath{\Conid{Pure}} constructor
by the compiler.

The language is shipped with simplifiers for the Boolean operators \ensuremath{\vee} and
\ensuremath{\mathrel{\wedge}}; as well as the $\IfThenElse{\cdot}{\cdot}{\cdot}$ operator and some numeric operators.
These simplifiers have great impact in temporal logics with references
to the future, where values can often be resolved at an instant with
the information retrieved up to that point---without the need to wait
until all future values are resolved.
To make specifications more readable, we use the Haskell extension
\ensuremath{\Conid{RebindableSyntax}} and define a function \ensuremath{\Conid{IfThenElse}} to be used every
time an $\IfThenElse{b}{e_0}{e_1}$ expression is used in a
specification:

{\mysize 
\begin{hscode}\SaveRestoreHook
\column{B}{@{}>{\hspre}l<{\hspost}@{}}%
\column{E}{@{}>{\hspre}l<{\hspost}@{}}%
\>[B]{}\Varid{ite}\mathbin{::}\Conid{Bool}\to \Varid{a}\to \Varid{a}\to \Varid{a}{}\<[E]%
\\
\>[B]{}\Varid{ite}\;\Conid{True}\;\Varid{x}\;\anonymous \mathrel{=}\Varid{x}{}\<[E]%
\\
\>[B]{}\Varid{ite}\;\Conid{False}\;\anonymous \;\Varid{y}\mathrel{=}\Varid{y}{}\<[E]%
\\[\blanklineskip]%
\>[B]{}\Varid{ifThenElse}\mathbin{::}\Conid{Typeable}\;\Varid{a}\Rightarrow \Conid{Expr}\;\Conid{Bool}\to \Conid{Expr}\;\Varid{a}\to \Conid{Expr}\;\Varid{a}\to \Conid{Expr}\;\Varid{a}{}\<[E]%
\\
\>[B]{}\Varid{ifThenElse}\;\Varid{b}\;\Varid{x}\;\Varid{y}\mathrel{=}\Varid{ite}~\boldsymbol{\mathop{\langle \$ \rangle}}~\Varid{b}~\boldsymbol{\mathop{\langle \star \rangle}}~\Varid{x}~\boldsymbol{\mathop{\langle \star \rangle}}~\Varid{y}{}\<[E]%
\ColumnHook
\end{hscode}\resethooks
}

This version of \ensuremath{\Varid{ifThenElse}} simply lifts the \ensuremath{\Varid{ifThenElse}} function in
the \ensuremath{\Conid{Prelude}}.
We show a simplifier with two simplifications of the
$\IfThenElse{b}{e_0}{e_1}$ which are implemented as a library of
\hlola.
Arguments in the simplifier are elements of the optional type \ensuremath{\Conid{Maybe}},
where
% an argument of
\ensuremath{\Conid{Nothing}} represents that the corresponding argument has not yet been
resolved, and
% an argument of
\ensuremath{\Conid{Just}\;\Varid{x}} represents that \ensuremath{\Varid{x}} is the argument for the function.

The value produced is also an optional element:
the returned value \ensuremath{\Conid{Nothing}} means that the arguments provided were
not enough to solve the function, while the returned value \ensuremath{\Conid{Just}\;\Varid{x}}
represents that the value of the function application is \ensuremath{\Varid{x}}
regardless of the missing arguments.

{\mysize 
\begin{hscode}\SaveRestoreHook
\column{B}{@{}>{\hspre}l<{\hspost}@{}}%
\column{E}{@{}>{\hspre}l<{\hspost}@{}}%
\>[B]{}\Varid{ite}\mathbin{::}\Conid{Maybe}\;\Conid{Bool}\to \Conid{Maybe}\;\Varid{a}\to \Conid{Maybe}\;\Varid{a}\to \Conid{Maybe}\;\Varid{a}{}\<[E]%
\\
\>[B]{}\Varid{ite}\;(\Conid{Just}\;\Conid{True})\;\Varid{x}\;\anonymous \mathrel{=}\Varid{x}{}\<[E]%
\\
\>[B]{}\Varid{ite}\;(\Conid{Just}\;\Conid{False})\;\anonymous \;\Varid{y}\mathrel{=}\Varid{y}{}\<[E]%
\\
\>[B]{}\Varid{ite}\;\Conid{Nothing}\;\anonymous \;\anonymous \mathrel{=}\Conid{Nothing}{}\<[E]%
\ColumnHook
\end{hscode}\resethooks

\ensuremath{\Varid{ifThenElse}\mathbin{::}\Conid{Typeable}\;\Varid{a}\Rightarrow \Conid{Expr}\;\Conid{Bool}\to \Conid{Expr}\;\Varid{a}\to \Conid{Expr}\;\Varid{a}\to \Conid{Expr}\;\Varid{a}}

\ensuremath{\Varid{ifThenElse}\;\Varid{b}\;\Varid{t}\;\Varid{e}\mathrel{=}\Varid{getsmp3}\;\Varid{ite}~\boldsymbol{\mathop{\langle \$ \rangle}}~\Varid{b}~\boldsymbol{\mathop{\langle \star \rangle}}~\Varid{t}~\boldsymbol{\mathop{\langle \star \rangle}}~\Varid{e}}
}

The auxiliary function \ensuremath{\Varid{getsmp3}} builds the \ensuremath{\Conid{Simplifier}} of a ternary
function.
% given its definition.
%
The case \ensuremath{\Varid{ite}\;(\Conid{Just}\;\Conid{True})\;\Varid{x}\;\anonymous \mathrel{=}\Varid{x}} implements the first simplification: if the
first argument has been resolved and its value is \ensuremath{\Conid{True}}, we return the value of
the second argument.
If the second argument has been resolved and thus its value is \ensuremath{\Conid{Just}\;\Varid{x}}, we simply return it.
If the second argument has not been resolved, then we cannot produce a
value, and hence return the same \ensuremath{\Conid{Nothing}}.
The second simplification is analogous in case the first
argument is \ensuremath{\Conid{False}}, but returning the third argument.
\subsubsection{Parameterized streams}
\label{sec:staticparam}
Static parametrization is a feature of some SRV systems which allows
instantiating an abstract specification.
This is useful to reuse repetitive specifications and capture the
essence of a stream definition, abstracting away the specific values.
Section~\ref{sec:libraries} shows how this feature is used to
concisely implement several monitoring languages as libraries in
\hlola.
This feature is implemented in Lola2.0~\cite{faymonville16stream} as
well as in TeSSLa~\cite{convent18tessla} using an ad-hoc macro feature
in the tool chain.
Here we show how static parametrization can be obtained directly using
Haskell features.
Consider again the specification of $\Once s$ shown in
Ex.~\ref{langdesign:once_s_spec}:

{\mysize
\ensuremath{\Varid{once\char95 s}\mathbin{::}\Conid{Stream}\;\Conid{Bool}}

\ensuremath{\Varid{once\char95 s}\mathrel{=}\text{\ttfamily \char34 once\char95 s\char34}\boldsymbol{=:}\Varid{once\char95 s}\boldsymbol{:\!\!@}(\mathbin{-}\mathrm{1},\Conid{False})\mathrel{\vee}\mathbf{Now}\;\Varid{s}}
}

If we want to define a stream to compute $\Once r$, we would have to
define a stream \ensuremath{\Varid{once\char95 r}} whose definition is almost identical to the
definition of \ensuremath{\Varid{once\char95 s}}.
This leads to code duplication and hard to maintain specifications.

Instead of defining an output stream \ensuremath{\Varid{once\char95 s}} specifically for \ensuremath{\Varid{s}}, we
aim to write a general stream \ensuremath{\Varid{once}} parameterized by a Boolean
stream.
We can use Haskell as a macro system to programmatically define
specifications, effectively implementing static parameterization.

\begin{example}
    \label{staticparam:once_spec}
The definition of \ensuremath{\Varid{once}} in \hlola using static parameterization is:

{\mysize
\ensuremath{\Varid{once}\mathbin{::}\Conid{Stream}\;\Conid{Bool}\to \Conid{Stream}\;\Conid{Bool}}

\ensuremath{\Varid{once}\;\Varid{s}\mathrel{=}\text{\ttfamily \char34 once\char34}\boldsymbol{<\!:}\Varid{s}\boldsymbol{=:}\Varid{once}\;\Varid{s}\boldsymbol{:\!\!@}(\mathbin{-}\mathrm{1},\Conid{False})\mathrel{\vee}\mathbf{Now}\;\Varid{s}}
}
\end{example}

\noindent Note that we simply abstracted away the occurrences of \ensuremath{\Varid{s}}.
To avoid name clashes among different instantiations of \ensuremath{\Varid{once}}, we
concatenate the string \ensuremath{\text{\ttfamily \char34 once\char34}} with the name of the argument stream
\ensuremath{\Varid{s}}, by using the operator \ensuremath{\boldsymbol{<\!:}}.
Static parametrization is used extensively to implement libraries as
described in the next section.
\section{Extensible libraries in HLola}
\label{sec:libraries}
One of the benefits of implementing an eDSL is that we can reuse the
library system of the host language to modularize and organize the
code.
The Haskell module system allows importing third parties libraries, as
well as developing new libraries;
\hlola ships with some predefined theories and stream-specific
libraries.
In this section we show an overview of the stream-specific libraries.

\subsubsection{Past-LTL.}
The operators of Past-LTL~\cite{benedetti03PastLTL} can be described
using the \lola specification language (e.g. $\Once$ from
Ex.~\ref{staticparam:once_spec}).
We now show the implementation of the Past-LTL constructors $\Since$
and $\LTLPrev$ as can be found in \ensuremath{\Conid{\Conid{Lib}.LTL}}:

\vspace{1em}
{\mysize
  \begin{tabular}{ll}
    \ensuremath{\Varid{since}} & \ensuremath{\mathbin{::}} \ensuremath{\Conid{Stream}\;\Conid{Bool}\to \Conid{Stream}\;\Conid{Bool}\to \Conid{Stream}\;\Conid{Bool}} \\
    \ensuremath{\Varid{yesterday}} & \ensuremath{\mathbin{::}} \ensuremath{\Conid{Stream}\;\Conid{Bool}\to \Conid{Stream}\;\Conid{Bool}} \\
    \multicolumn{2}{l}{
    \begin{tabular}{@{\hspace{-1.8em}}l@{}l}
\begin{minipage}[c]{0.6\textwidth}\footnotesize
\begin{hscode}\SaveRestoreHook
\column{B}{@{}>{\hspre}l<{\hspost}@{}}%
\column{3}{@{}>{\hspre}l<{\hspost}@{}}%
\column{5}{@{}>{\hspre}l<{\hspost}@{}}%
\column{11}{@{}>{\hspre}l<{\hspost}@{}}%
\column{E}{@{}>{\hspre}l<{\hspost}@{}}%
\>[B]{}\Varid{since}\;\Varid{p}\;\Varid{q}\mathrel{=}\mathbf{let}{}\<[E]%
\\
\>[B]{}\hsindent{5}{}\<[5]%
\>[5]{}\Varid{name}\mathrel{=}\text{\ttfamily \char34 since\char34}\boldsymbol{<\!:}\Varid{p}\boldsymbol{<\!:}\Varid{q}{}\<[E]%
\\
\>[B]{}\hsindent{5}{}\<[5]%
\>[5]{}\Varid{body}\mathrel{=}\mathbf{Now}\;\Varid{q}\mathrel{\vee}{}\<[E]%
\\
\>[5]{}\hsindent{6}{}\<[11]%
\>[11]{}\qquad(\mathbf{Now}\;\Varid{p}\mathrel{\wedge}\Varid{p}\mathbin{`\Varid{since}`}\Varid{q}\boldsymbol{:\!\!@}(\mathbin{-}\mathrm{1},\Conid{False})){}\<[E]%
\\
\>[B]{}\hsindent{3}{}\<[3]%
\>[3]{}\mathbf{in}\;\Varid{name}\boldsymbol{=:}\Varid{body}{}\<[E]%
\ColumnHook
\end{hscode}\resethooks
\end{minipage}
  &
\begin{minipage}[c]{0.4\textwidth}\footnotesize
  \begin{hscode}\SaveRestoreHook
\column{B}{@{}>{\hspre}l<{\hspost}@{}}%
\column{5}{@{}>{\hspre}l<{\hspost}@{}}%
\column{7}{@{}>{\hspre}l<{\hspost}@{}}%
\column{9}{@{}>{\hspre}l<{\hspost}@{}}%
\column{E}{@{}>{\hspre}l<{\hspost}@{}}%
\>[5]{}\Varid{yesterday}\;\Varid{p}\mathrel{=}\mathbf{let}{}\<[E]%
\\
\>[5]{}\hsindent{4}{}\<[9]%
\>[9]{}\Varid{name}\mathrel{=}\text{\ttfamily \char34 yesterday\char34}\boldsymbol{<\!:}\Varid{p}{}\<[E]%
\\
\>[5]{}\hsindent{4}{}\<[9]%
\>[9]{}\Varid{body}\mathrel{=}\Varid{p}\boldsymbol{:\!\!@}(\mathbin{-}\mathrm{1},\Conid{False}){}\<[E]%
\\
\>[5]{}\hsindent{2}{}\<[7]%
\>[7]{}\mathbf{in}\;\Varid{name}\boldsymbol{=:}\Varid{body}{}\<[E]%
\ColumnHook
\end{hscode}\resethooks
%\vspace{2em}
\end{minipage}
    \end{tabular}}
  \end{tabular}   
}

Given two Boolean streams \ensuremath{\Varid{p}} and \ensuremath{\Varid{q}}, the Boolean stream \ensuremath{\Varid{p}\mathbin{`\Varid{since}`}\Varid{q}} is \ensuremath{\Conid{True}} if \ensuremath{\Varid{q}} has ever been \ensuremath{\Conid{True}}, and \ensuremath{\Varid{p}} has been \ensuremath{\Conid{True}} since the last
time \ensuremath{\Varid{q}} became \ensuremath{\Conid{True}}.
One can simply \ensuremath{\mathbf{import}\;\Conid{\Conid{Lib}.LTL}} and then define streams like:
\ensuremath{\Varid{property}\mathrel{=}\Varid{yesterday}\;(\Varid{p}\mathbin{`\Varid{since}`}\Varid{q})}.
\begin{example}
We show an example of a Past-LTL property for a sender/receiver model
taken from~\cite{benedetti03PastLTL}:
\(
  \Always (\ensuremath{\Varid{snd}}.\ensuremath{\Varid{state}} = \ensuremath{\Varid{waitForAck}} \rightarrow \LTLPrev \HasAlwaysBeen
  \ensuremath{\Varid{snd}}.\ensuremath{\Varid{state}} \neq \ensuremath{\Varid{waitForAck}})
\).

\noindent Using \hlola, we define a type to represent the possible
states of the sender, deriving a \ensuremath{\Conid{FromJSON}} instance to use it as the
type of an input stream \ensuremath{\Varid{sndrState}}:

\vspace{0.5em}
{\mysize
  \noindent
\ensuremath{\mathbf{data}\;\Conid{SndrState}\mathrel{=}\Conid{Get}\mid \Conid{Send}\mid \Conid{WaitForAck}\;\mathbf{deriving}\;(\Conid{Generic},\Conid{Read},\Conid{FromJSON},\Conid{Eq})}
}

\noindent Then, we define the property as a Boolean output stream:

{\mysize \begin{hscode}\SaveRestoreHook
\column{B}{@{}>{\hspre}l<{\hspost}@{}}%
\column{E}{@{}>{\hspre}l<{\hspost}@{}}%
\>[B]{}\Varid{sndrState}\mathbin{::}\Conid{Stream}\;\Conid{SndrState}{}\<[E]%
\\
\>[B]{}\Varid{sndrState}\mathrel{=}\mathbf{Input}\;\text{\ttfamily \char34 senderState\char34}{}\<[E]%
\\[\blanklineskip]%
\>[B]{}\Varid{sndrNotWaiting}\mathbin{::}\Conid{Stream}\;\Conid{Bool}{}\<[E]%
\\
\>[B]{}\Varid{sndrNotWaiting}\mathrel{=}\text{\ttfamily \char34 sndrNotWaiting\char34}\boldsymbol{=:}\mathbf{Now}\;\Varid{sndrState}\mathbin{/==}\mathbf{Leaf}\;\Conid{WaitForAck}{}\<[E]%
\\[\blanklineskip]%
\>[B]{}\Varid{prop}\mathbin{::}\Conid{Stream}\;\Conid{Bool}{}\<[E]%
\\
\>[B]{}\Varid{prop}\mathrel{=}\mathbf{let}\;\Varid{sndrWaitingAck}\mathrel{=}\mathbf{Now}\;\Varid{sndrState}\mathbin{===}\mathbf{Leaf}\;\Conid{WaitForAck}{}\<[E]%
\\
\>[B]{}\qquad\qquad\quad\;\Varid{startedWaiting}\mathrel{=}\Varid{yesterday}\;(\Varid{historically}\;\Varid{sndrNotWaiting}){}\<[E]%
\\
\>[B]{}\qquad\quad\;\mathbf{in}\;\text{\ttfamily \char34 prop\char34}\boldsymbol{=:}\Varid{sndrWaitingAck}\mathbin{`\Varid{implies}`}\mathbf{Now}\;\Varid{startedWaiting}{}\<[E]%
\ColumnHook
\end{hscode}\resethooks
}
\end{example}

\vspace{-1.5em}
\subsubsection{MTL.}

%\vspace{-1em}
%
Metric Temporal Logic \cite{koymans90specifying} is an extension of
LTL with time constraints that give upper and lower bounds on the
temporal intervals.
We show the implementation of the MTL constructors $\UNTIL$ and
$\Event$ (the definitions of the \ensuremath{\Varid{name}} associated with the streams
have been omitted for clarity):

 {\mysize
 \begin{hscode}\SaveRestoreHook
\column{B}{@{}>{\hspre}l<{\hspost}@{}}%
\column{3}{@{}>{\hspre}l<{\hspost}@{}}%
\column{5}{@{}>{\hspre}l<{\hspost}@{}}%
\column{E}{@{}>{\hspre}l<{\hspost}@{}}%
\>[B]{}\!\!\Varid{until}\mathbin{::}(\Conid{Int},\Conid{Int})\to \Conid{Stream}\;\Conid{Bool}\to \Conid{Stream}\;\Conid{Bool}\to \Conid{Stream}\;\Conid{Bool}{}\<[E]%
\\
\>[B]{}\!\!\Varid{until}\;(\Varid{a},\Varid{b})\;\Varid{p}\;\Varid{q}\mathrel{=}\Varid{name}\boldsymbol{=:}\Varid{until'}\;\Varid{a}\;\Varid{b}\;\Varid{p}\;\Varid{q}{}\<[E]%
\\
\>[B]{}\hsindent{3}{}\<[3]%
\>[3]{}\mathbf{where}{}\<[E]%
\\
\>[B]{}\hsindent{3}{}\<[3]%
\>[3]{}\Varid{until'}\;\Varid{a}\;\Varid{b}\;\Varid{p}\;\Varid{q}{}\<[E]%
\\
\>[3]{}\hsindent{2}{}\<[5]%
\>[5]{}\mid \Varid{a}\equiv \Varid{b}\mathrel{=}\Varid{q}\boldsymbol{:\!\!@}(\Varid{b},\Conid{False}){}\<[E]%
\\
\>[3]{}\hsindent{2}{}\<[5]%
\>[5]{}\mid \Varid{otherwise}\mathrel{=}(\Varid{q}\boldsymbol{:\!\!@}(\Varid{a},\Conid{False})\mathrel{\vee}(\Varid{p}\boldsymbol{:\!\!@}(\Varid{a},\Conid{True})\mathrel{\wedge}\Varid{until'}\;(\Varid{a}\mathbin{+}\mathrm{1})\;\Varid{b}\;\Varid{p}\;\Varid{q})){}\<[E]%
\\[\blanklineskip]%
\>[B]{}\!\!\Varid{eventually}\mathbin{::}(\Conid{Int},\Conid{Int})\to \Conid{Stream}\;\Conid{Bool}\to \Conid{Stream}\;\Conid{Bool}{}\<[E]%
\\
\>[B]{}\!\!\Varid{eventually}\;(\Varid{a},\Varid{b})\;\Varid{p}\mathrel{=}\Varid{name}\boldsymbol{=:}\Varid{foldl}\;(\mathrel{\vee})\;(\mathbf{Leaf}\;\Conid{False})\;[\mskip1.5mu \Varid{p}\boldsymbol{:\!\!@}(\Varid{i},\Conid{False})\mid \Varid{i}\leftarrow [\mskip1.5mu \Varid{a}\mathinner{\ldotp\ldotp}\Varid{b}\mskip1.5mu]\mskip1.5mu]{}\<[E]%
\ColumnHook
\end{hscode}\resethooks
}
\noindent The stream \ensuremath{\Varid{until}} is parameterized by two integers,
which are the boundaries of the interval, and two Boolean streams to
model the formula $p \UNTIL_{[a,b]} q$.
We use recursion to programmatically define the \expression of
\ensuremath{\Varid{until}}, which will be unfolded at compile time for the dependency
graph sanity check.
This expansion can
\begin{wrapfigure}[6]{c}{0.30\textwidth}
  \vspace{-2.1em}
    \includegraphics[scale=0.85]{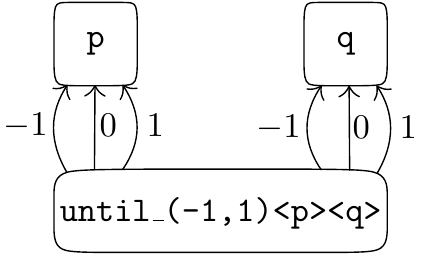}
\end{wrapfigure}
be observed in the dependency graph of a specification that uses \ensuremath{\Varid{until}}, for
example, \ensuremath{\Varid{property}\mathrel{=}\Varid{until}\;(\mathbin{-}\mathrm{1},\mathrm{1})\;\Varid{p}\;\Varid{q}}, which checks that a stream \ensuremath{\Varid{p}}
is \ensuremath{\Conid{True}} until \ensuremath{\Varid{q}} is \ensuremath{\Conid{True}} in the interval \ensuremath{(\mathbin{-}\mathrm{1},\mathrm{1})}, which is shown
on the right.
\newcommand{\mission}{\textit{mission}}

In~\cite{reinbacher14MTLTL}, Reinbacher et al. introduce Mission-Time
LTL, a projection of LTL for systems which are bounded to a certain
mission time.
They propose a translation of each LTL operator to its corresponding
MTL operator, using $[0,\mission_t]$ as the temporal interval, where
$\mission_t$ represents how the duration of the mission.
The ability of \hlola to monitor MTL can be used to monitor
Mission-Time LTL through this translation.
\begin{example}
\label{ex:mitl}
We show an example of an MTL property taken
from~\cite{RecentMTLResults}:
\(
\Always (alarm \rightarrow (\Event_{[0,10]} allclear \lor
\Event_{[10,10]} shutdown))
\)

\noindent This property uses MTL to establish deadlines between environment
events and the corresponding system responses.
In particular, the property assesses that an \emph{alarm} is followed
by a \emph{shutdown} event in exactly $10$ time units unless \emph{all
clear} is sounded first.
We consider three Boolean input streams \ensuremath{\Varid{alarm}}, \ensuremath{\Varid{allclear}} and
\ensuremath{\Varid{shutdown}}---which indicate if the corresponding event is detected---and
define an output stream that captures whether the property holds:

{\mysize\begin{hscode}\SaveRestoreHook
\column{B}{@{}>{\hspre}l<{\hspost}@{}}%
\column{3}{@{}>{\hspre}l<{\hspost}@{}}%
\column{14}{@{}>{\hspre}l<{\hspost}@{}}%
\column{E}{@{}>{\hspre}l<{\hspost}@{}}%
\>[B]{}\Varid{alarm}\mathrel{=}\mathbf{Input}\;\text{\ttfamily \char34 alarm\char34}\mathbin{::}\Conid{Stream}\;\Conid{Bool}{}\<[E]%
\\
\>[B]{}\Varid{allclear}\mathrel{=}\mathbf{Input}\;\text{\ttfamily \char34 allclear\char34}\mathbin{::}\Conid{Stream}\;\Conid{Bool}{}\<[E]%
\\
\>[B]{}\Varid{shutdown}\mathrel{=}\mathbf{Input}\;\text{\ttfamily \char34 shutdown\char34}\mathbin{::}\Conid{Stream}\;\Conid{Bool}{}\<[E]%
\\[\blanklineskip]%
\>[B]{}\Varid{prop}\mathbin{::}\Conid{Stream}\;\Conid{Bool}{}\<[E]%
\\
\>[B]{}\Varid{prop}\mathrel{=}\text{\ttfamily \char34 prop\char34}\boldsymbol{=:}\mathbf{Now}\;\Varid{alarm}\mathbin{`\Varid{implies}`}\mathbf{Now}\;\Varid{willClear}\mathrel{\vee}\mathbf{Now}\;\Varid{willShutdown}{}\<[E]%
\\
\>[B]{}\hsindent{3}{}\<[3]%
\>[3]{}\mathbf{where}\;\Varid{willClear}\mathrel{=}\Varid{eventually}\;(\mathrm{0},\mathrm{10})\;\Varid{allclear}{}\<[E]%
\\
\>[B]{}\hsindent{3}{}\<[3]%
\>[3]{}\qquad\quad\;{}\<[14]%
\>[14]{}\Varid{willShutdown}\mathrel{=}\Varid{eventually}\;(\mathrm{10},\mathrm{10})\;\Varid{shutdown}{}\<[E]%
\ColumnHook
\end{hscode}\resethooks
}
\end{example}
\section{Implementation and Empirical evaluation}
\label{sec:empirical}

\newcommand{\EngineLOC}{
\begin{tabular}{|l|c|}
  \multicolumn{2}{c}{Engine} \\
  \hline
  Files: Engine/ & LoC  \\
  \hline
  Engine.hs & 176 \\
  Focus.hs & 39 \\
  & \\ & \\ \hline
  \textbf{Total} & 215 \\ \hline
\end{tabular}
}
\newcommand{\SyntaxLOC}{
\addvbuffer[0.0cm 0.0cm]{
  \begin{tabular}{|l|c|}
  \multicolumn{2}{c}{Syntax} \\
  \hline
  Files: Syntax/ & LoC \\
  \hline
  Booleans.hs & 37  \\
  HLPrelude.hs \phantom{aa} & 3 \\
  Num.hs & 26 \\
  Ord.hs & 18 \\ \hline
  \textbf{Total} & 102 \\ \hline
  \end{tabular}
}
}
\newcommand{\LibrariesLOC}{
\begin{tabular}{|l|c|}
  \multicolumn{2}{c}{Libraries} \\
  \hline
  Files: Lib/ & LoC \\\hline
  LTL.hs & 21  \\
  MTL.hs & 29 \\
  Pinescript.hs & 41\\
  Utils.hs & 13\\ \hline
  \textbf{Total} & 104 \\ \hline
\end{tabular}
}
\newcommand{\LanguageLOC}{
\begin{tabular}{|l|c|}
  \multicolumn{2}{c}{Language and input} \\
  \hline
  Files: ./\phantom{a} & LoC \\ \hline
  DecDyn.hs & 87  \\
  InFromFile.hs & 51 \\
  Lola.hs & 62 \\
  StaticAnalysis.hs & 78 \\ \hline
  \textbf{Total} & 278 \\   \hline
\end{tabular}
}

\newcommand{\LanguageWFLOC}{
\addvbuffer[0.0em -0.005em]{
  \begin{tabular}{cc}
    \LanguageLOC & \LanguageLOC
  \end{tabular}
}
}

The implementation of \hlola
requires no code for the parser and type checker, since it reuses
those from the Haskell compiler.
The table
below shows the number of lines for the full \hlola implementation.
 
{\small
\noindent\begin{tabular}[t]{@{}c@{}c@{}c@{}c@{}}
  \LanguageLOC &
  \EngineLOC &
  \SyntaxLOC &
  \LibrariesLOC
\end{tabular}  
}

In summary, the core of the tool has $493$ lines, while the utils account for
$206$ lines, giving a total of $699$ lines.
This compares to the tens of thousands of lines of a parser and
runtime system of a typical stand-alone tool.
In the rest of this section we summarize how using Haskell enables the
use of available tools, and then report on an empirical evaluation of
\hlola.

\subsubsection{Haskell tools.}
The use of Haskell as a host language allows us to use existing tools
to improve \hlola specifications, such as LiquidHaskell and
QuickCheck.

Liquid Haskell~\cite{vazou14liquid} enriches the type system with
refinement types that allow more precise descriptions of the types of
the elements in a Haskell program.
In our case we can use Liquid Haskell to express specifications with
more precision.
For example, we can prevent a specification that adds the last $n$
elements from being used with a negative $n$:

{\mysize 
% \begin{code}
\ensuremath{\mbox{\commentbegin \text{\ttfamily ~nsum~\char58{}\char58{}~Stream~Int~\char45{}\char62{}~Nat~\char45{}\char62{}~Stream~Int~} \commentend}}

\ensuremath{\Varid{nsum}\mathbin{::}\Conid{Stream}\;\Conid{Int}\to \Conid{Int}\to \Conid{Stream}\;\Conid{Int}}

\ensuremath{\Varid{nsum}\;\Varid{s}\;\Varid{n}\mathrel{=}\text{\ttfamily \char34 n\char95 sum\char34}\boldsymbol{<\!:}\Varid{s}\boldsymbol{<\!:}\Varid{n}\boldsymbol{=:}\Varid{nsum}\;\Varid{s}\;\Varid{n}\boldsymbol{:\!\!@}(\mathbin{-}\mathrm{1},\mathrm{0})\mathbin{+}\mathbf{Now}\;\Varid{s}\mathbin{-}\Varid{s}\boldsymbol{:\!\!@}(\mathbin{-}\Varid{n},\mathrm{0})}
% \end{code}
}

\noindent Then, given a stream $r$ of type \ensuremath{\Conid{Stream}\;\Conid{Int}} we can attempt
to define a stream \ensuremath{\Varid{s}} that computes the sum of the last $5$ values on
stream \ensuremath{\Varid{r}} as \ensuremath{\Varid{s}\mathrel{=}\Varid{nsum}\;\Varid{r}\;\mathrm{5}}.
Running LiquidHaskell with \text{\ttfamily \char45{}\char45{}no\char45{}termination} allows the
recursive definition of \ensuremath{\Varid{n}} over this specification, which yields no
error, but running LiquidHaskell on \ensuremath{\Varid{s'}\mathrel{=}\Varid{nsum}\;\Varid{r}\;(\mathbin{-}\mathrm{1})} produces a typing
error.
 
% \paragraph{QuickCheck.}
%

\begin{figure}[b!]
\centering
%\begin{tabular}{|l|l|c|c|c|c|c|c|c|}
\scalebox{0.7}{\begin{tabular}{ccc}
  \includegraphics[scale=0.7]{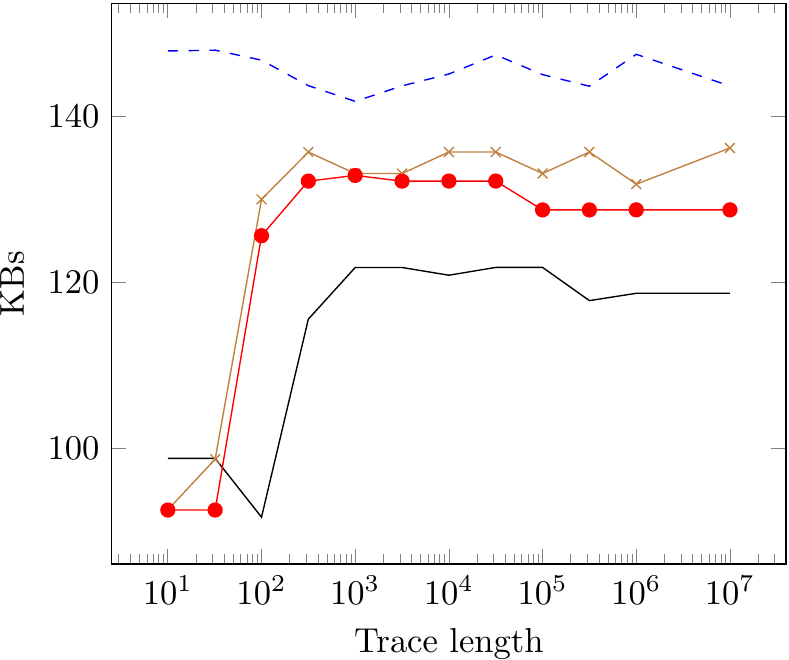} &
  \includegraphics[scale=0.7]{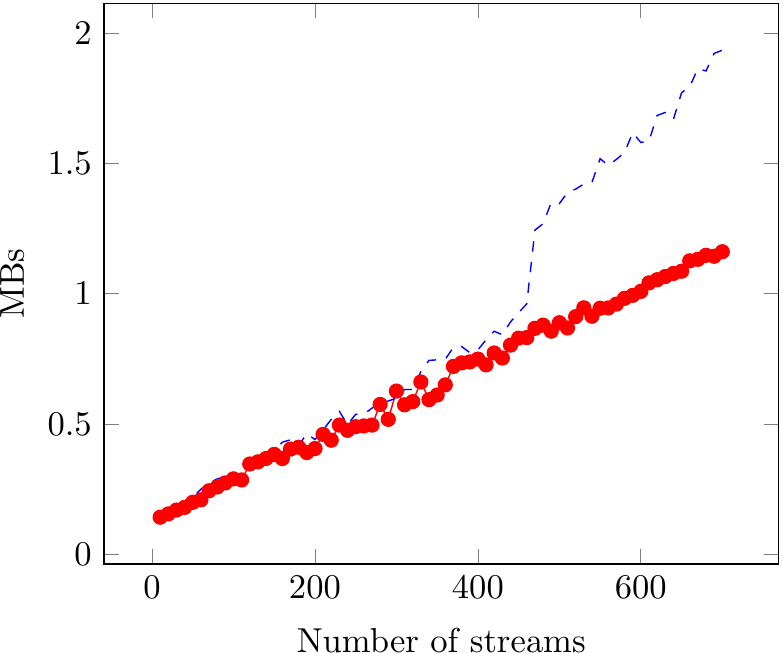} &
  \includegraphics[scale=0.7]{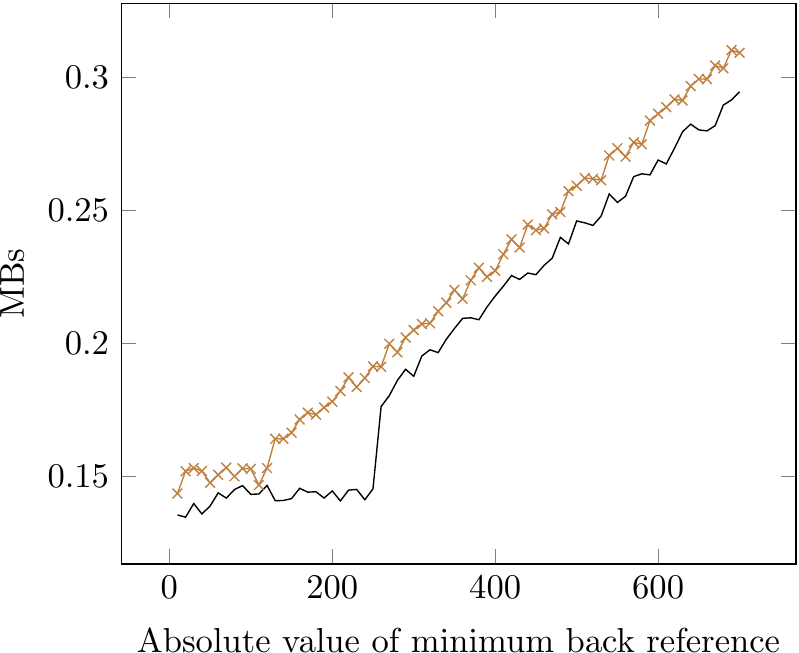} \\
  (a) Memory wrt trace length &
  (b)   Memory wrt number of streams &
  (c) Memory wrt \minBackRef
\end{tabular}}
\caption{Empirical evaluation}
\label{tab:empirical:plots}
\end{figure}

QuickCheck~\cite{claessen00quickcheck} is a tool to perform random
testing of Haskell programs, which we can easily use for \hlola
specifications.
For example, we can assess that the first instant at which a Boolean
stream \ensuremath{\Varid{p}} is \ensuremath{\Conid{False}} is exactly one instant after the last instant at
which $\HasAlwaysBeen p$ is \ensuremath{\Conid{True}}, increasing our confidence on the
implementation of the Past-LTL $\HasAlwaysBeen$ operator, as follows:

{\mysize \begin{hscode}\SaveRestoreHook
\column{B}{@{}>{\hspre}l<{\hspost}@{}}%
\column{3}{@{}>{\hspre}l<{\hspost}@{}}%
\column{5}{@{}>{\hspre}l<{\hspost}@{}}%
\column{7}{@{}>{\hspre}l<{\hspost}@{}}%
\column{E}{@{}>{\hspre}l<{\hspost}@{}}%
\>[B]{}\Varid{main}\mathbin{::}\Conid{IO}\;(){}\<[E]%
\\
\>[B]{}\Varid{main}\mathrel{=}\Varid{quickCheck}\;\Varid{historicallyIsCorrect}{}\<[E]%
\\[\blanklineskip]%
\>[B]{}\Varid{historicallyIsCorrect}\mathbin{::}[\mskip1.5mu \Conid{Bool}\mskip1.5mu]\to \Conid{Bool}{}\<[E]%
\\
\>[B]{}\Varid{historicallyIsCorrect}\;\Varid{bs}\mathrel{=}\mathbf{let}{}\<[E]%
\\
\>[B]{}\hsindent{5}{}\<[5]%
\>[5]{}\Varid{p}\mathrel{=}\mathbf{Input}\;\text{\ttfamily \char34 p\char34}{}\<[E]%
\\
\>[B]{}\hsindent{5}{}\<[5]%
\>[5]{}\Varid{spec}\mathrel{=}[\mskip1.5mu \Varid{out}\;(\Varid{historically}\;\Varid{p}),\Varid{out}\;\Varid{p}\mskip1.5mu]{}\<[E]%
\\
\>[B]{}\hsindent{5}{}\<[5]%
\>[5]{}\Varid{trace}\mathrel{=}\Varid{runLib}\;\Varid{spec}\;(\Varid{map}\;(\Varid{singleton}\;\text{\ttfamily \char34 p\char34}\mathbin{\circ}\Varid{toDyn})\;\Varid{bs}){}\<[E]%
\\
\>[B]{}\hsindent{5}{}\<[5]%
\>[5]{}\Varid{mMinNotP}\mathrel{=}\Varid{findIndex}\;\neg \;(\Varid{map}\;(\Varid{retrieve}\;\text{\ttfamily \char34 p\char34})\;\Varid{trace}){}\<[E]%
\\
\>[B]{}\hsindent{5}{}\<[5]%
\>[5]{}\Varid{mMaxHistP}\mathrel{=}\Varid{findLastIndex}\;\Varid{id}{}\<[E]%
\\
\>[5]{}\hsindent{2}{}\<[7]%
\>[7]{}(\Varid{map}\;(\Varid{retrieve}\;\text{\ttfamily \char34 historically<p>\char34})\;\Varid{trace}){}\<[E]%
\\
\>[B]{}\hsindent{5}{}\<[5]%
\>[5]{}\Varid{minNotP}\mathrel{=}\Varid{fromMaybe}\;(\Varid{length}\;\Varid{trace})\;\Varid{mMinNotP}{}\<[E]%
\\
\>[B]{}\hsindent{5}{}\<[5]%
\>[5]{}\Varid{maxHistP}\mathrel{=}\Varid{fromMaybe}\;(\mathbin{-}\mathrm{1})\;\Varid{mMaxHistP}{}\<[E]%
\\
\>[B]{}\hsindent{3}{}\<[3]%
\>[3]{}\mathbf{in}\;\Varid{minNotP}\mathbin{-}\mathrm{1}\equiv \Varid{maxHistP}{}\<[E]%
\\
\>[B]{}\hsindent{3}{}\<[3]%
\>[3]{}\mathbf{where}{}\<[E]%
\\
\>[3]{}\hsindent{2}{}\<[5]%
\>[5]{}\Varid{retrieve}\;\Varid{strid}\;\Varid{m}\mathrel{=}\Varid{fromJust}\mathbin{\circ}\Varid{fromDynamic}\;(\Varid{m}\mathbin{!}\Varid{strid}){}\<[E]%
\\
\>[3]{}\hsindent{2}{}\<[5]%
\>[5]{}\Varid{findLastIndex}\;\Varid{f}\;\Varid{l}\mathrel{=}\Varid{findIndex}\;\Varid{f}\;(\Varid{reverse}\;\Varid{l})\bind {}\<[E]%
\\
\>[5]{}\hsindent{2}{}\<[7]%
\>[7]{}\lambda \Varid{a}\to \Conid{Just}\;(\Varid{length}\;\Varid{l}\mathbin{-}\Varid{a}\mathbin{-}\mathrm{1}){}\<[E]%
\ColumnHook
\end{hscode}\resethooks
}

%\vspace{-3em}
\subsubsection{Empirical evaluation.}
We report now on an empirical evaluation performed to assess whether
the engine behaves as theoretically expected in terms of memory usage.
The hardware platform over which the experiments were run is a MacBook
Pro with MacOS Catalina Version 10.15.4, with an Intel Core i5 at 2,5
GHz and 8 GB of RAM.

The first two \declarations calculate if an input \ensuremath{\Conid{Boolean}} stream \ensuremath{\Varid{p}}
is periodic with period \ensuremath{\Varid{n}}.
This is a simple, yet interesting property to assess in embedded
systems.
We specify this property in two different ways.
In the first \declaration, we define a single stream which compares
the current value of \ensuremath{\Varid{p}} with its value \ensuremath{\Varid{n}} instants before:

{\mysize
  % \begin{code}
\ensuremath{\Varid{booleanPeriodWidth}\mathbin{::}\Conid{Int}\to \Conid{Stream}\;\Conid{Bool}}

\ensuremath{\Varid{booleanPeriodWidth}\;\Varid{n}\mathrel{=}\text{\ttfamily \char34 periodic\char95 width\char34}\boldsymbol{=:}\mathbf{Now}\;\Varid{p}\mathbin{===}\Varid{p}\boldsymbol{:\!\!@}(\mathbin{-}\Varid{n},\mathbf{Now}\;\Varid{p})}
% \end{code}
}
The data of this experiment is represented by the solid, unmarked
black curves in Fig.~\ref{tab:empirical:plots} (a) and (c).

In the second \declaration, we programmatically create $n+1$ streams
\ensuremath{\Varid{carrier}\;\Varid{i}}, with $i=0 \ldots n$ defined as a function that compares
its argument with the value of \ensuremath{\Varid{p}} \ensuremath{\Varid{i}} instants before, which is bound
by the partially applied equality function:
{\mysize
  \begin{hscode}\SaveRestoreHook
\column{B}{@{}>{\hspre}l<{\hspost}@{}}%
\column{9}{@{}>{\hspre}l<{\hspost}@{}}%
\column{E}{@{}>{\hspre}l<{\hspost}@{}}%
\>[B]{}\!\!\!\!\!\!\!\!\Varid{booleanPeriodHeight}\mathbin{::}\Conid{Int}\to \Conid{Stream}\;\Conid{Bool}{}\<[E]%
\\
\>[B]{}\!\!\!\!\!\!\!\!\Varid{booleanPeriodHeight}\;\Varid{n}\mathrel{=}\text{\ttfamily \char34 periodic\char95 height\char34}\boldsymbol{=:}\mathbf{Now}\;(\Varid{carrier}\;\Varid{n})~\boldsymbol{\mathop{\langle \star \rangle}}~\mathbf{Now}\;\Varid{p}{}\<[E]%
\\
\>[B]{}\!\!\!\!\mathbf{where}{}\<[E]%
\\
\>[B]{}\!\!\!\!{}\<[9]%
\>[9]{}\Varid{carrier}\;\mathrm{0}\mathrel{=}\text{\ttfamily \char34 carrier\char95 prd\char34}\boldsymbol{<\!:}\mathrm{0}\boldsymbol{=:}(\equiv )~\boldsymbol{\mathop{\langle \$ \rangle}}~\mathbf{Now}\;\Varid{p}{}\<[E]%
\\
\>[B]{}\!\!\!\!{}\<[9]%
\>[9]{}\Varid{carrier}\;\Varid{n}\mathrel{=}\text{\ttfamily \char34 carrier\char95 prd\char34}\boldsymbol{<\!:}\Varid{n}\boldsymbol{=:}\Varid{carrier}\;(\Varid{n}\mathbin{-}\mathrm{1})\boldsymbol{:\!\!@}(\mathbin{-}\mathrm{1},\mathbf{Leaf}\;(\Varid{const}\;\Conid{True})){}\<[E]%
\ColumnHook
\end{hscode}\resethooks
}
The data of this experiment is represented by the solid, circle-marked
red curves in Fig.~\ref{tab:empirical:plots} (a) and (b).

We also run a quantitative version of this \ensuremath{\Varid{n}}-period checker, whose
value is \ensuremath{\mathrm{100}} at a given instant if \ensuremath{\Varid{p}} at that instant is equal to
the value of \ensuremath{\Varid{p}} \ensuremath{\Varid{n}} instants ago; \ensuremath{\mathrm{50}} if it is equal to the value of
\ensuremath{\Varid{p}} \ensuremath{\Varid{n}\mathbin{-}\mathrm{1}} or \ensuremath{\Varid{n}\mathbin{+}\mathrm{1}} instants ago; \ensuremath{\mathrm{25}} if it is equal to the value of
\ensuremath{\Varid{p}} \ensuremath{\Varid{n}\mathbin{-}\mathrm{2}} or \ensuremath{\Varid{n}\mathbin{+}\mathrm{2}} instants ago; and \ensuremath{\mathrm{0}} otherwise:
Note that this specification has a value closer to \ensuremath{\mathrm{100}} when the
specification is closer to being periodic, and closer to \ensuremath{\mathrm{0}} when the
specification is further from being periodic.
This example illustrates how \hlola can be used to define quantitative
semantics of temporal logics, which is an active area of research in
Runtime Verification.
In this case we also define a version with a single stream
(represented by the solid, cross-marked brown curves in
Fig.~\ref{tab:empirical:plots} (a) and (c)), and a version with
auxiliary streams, each of which has an offset of \ensuremath{\mathbin{-}\mathrm{1}} at most
(represented by the dashed blue curves in
Fig.~\ref{tab:empirical:plots} (a) and (b)):

{\mysize \begin{hscode}\SaveRestoreHook
\column{B}{@{}>{\hspre}l<{\hspost}@{}}%
\column{3}{@{}>{\hspre}l<{\hspost}@{}}%
\column{13}{@{}>{\hspre}l<{\hspost}@{}}%
\column{E}{@{}>{\hspre}l<{\hspost}@{}}%
\>[B]{}\Varid{smoothPeriodWidth}\mathbin{::}\Conid{Int}\to \Conid{Stream}\;\Conid{Int}{}\<[E]%
\\
\>[B]{}\Varid{smoothPeriodWidth}\;\Varid{n}\mathrel{=}\text{\ttfamily \char34 smooth\char95 period\char95 width\char34}\boldsymbol{=:}{}\<[E]%
\\
\>[B]{}\hsindent{3}{}\<[3]%
\>[3]{}\mathbf{if}\;\mathbf{Now}\;\Varid{p}\mathbin{===}\Varid{p}\boldsymbol{:\!\!@}(\mathbin{-}\Varid{n},\mathbf{Now}\;\Varid{p})\;\mathbf{then}\;\mathrm{100}\;\mathbf{else}{}\<[E]%
\\
\>[B]{}\hsindent{3}{}\<[3]%
\>[3]{}\mathbf{if}\;\mathbf{Now}\;\Varid{p}\mathbin{===}\Varid{p}\boldsymbol{:\!\!@}(\mathbin{-}\Varid{n}\mathbin{-}\mathrm{1},\mathbf{Now}\;\Varid{p})\mathrel{\vee}{}\<[E]%
\\
\>[3]{}\hsindent{10}{}\<[13]%
\>[13]{}\mathbf{Now}\;\Varid{p}\mathbin{===}\Varid{p}\boldsymbol{:\!\!@}(\mathbin{-}\Varid{n}\mathbin{+}\mathrm{1},\mathbf{Now}\;\Varid{p})\;\mathbf{then}\;\mathrm{50}\;\mathbf{else}{}\<[E]%
\\
\>[B]{}\hsindent{3}{}\<[3]%
\>[3]{}\mathbf{if}\;\mathbf{Now}\;\Varid{p}\mathbin{===}\Varid{p}\boldsymbol{:\!\!@}(\mathbin{-}\Varid{n}\mathbin{-}\mathrm{2},\mathbf{Now}\;\Varid{p})\mathrel{\vee}{}\<[E]%
\\
\>[3]{}\hsindent{10}{}\<[13]%
\>[13]{}\mathbf{Now}\;\Varid{p}\mathbin{===}\Varid{p}\boldsymbol{:\!\!@}(\mathbin{-}\Varid{n}\mathbin{+}\mathrm{2},\mathbf{Now}\;\Varid{p})\;\mathbf{then}\;\mathrm{25}\;\mathbf{else}\;\mathrm{0}{}\<[E]%
\ColumnHook
\end{hscode}\resethooks
}
\vspace{-2em}
{\mysize \begin{hscode}\SaveRestoreHook
\column{B}{@{}>{\hspre}l<{\hspost}@{}}%
\column{3}{@{}>{\hspre}l<{\hspost}@{}}%
\column{6}{@{}>{\hspre}l<{\hspost}@{}}%
\column{9}{@{}>{\hspre}l<{\hspost}@{}}%
\column{E}{@{}>{\hspre}l<{\hspost}@{}}%
\>[B]{}\Varid{smoothPeriodHeight}\mathbin{::}\Conid{Int}\to \Conid{Stream}\;\Conid{Int}{}\<[E]%
\\
\>[B]{}\Varid{smoothPeriodHeight}\;\Varid{n}\mathrel{=}\text{\ttfamily \char34 smooth\char95 period\char95 height\char34}\boldsymbol{=:}{}\<[E]%
\\
\>[B]{}\hsindent{3}{}\<[3]%
\>[3]{}\mathbf{if}\;\mathbf{Now}\;(\Varid{carrier}\;\Varid{n})~\boldsymbol{\mathop{\langle \star \rangle}}~\mathbf{Now}\;\Varid{p}\;\mathbf{then}\;\mathrm{100}\;\mathbf{else}{}\<[E]%
\\
\>[B]{}\hsindent{3}{}\<[3]%
\>[3]{}\mathbf{if}\;\mathbf{Now}\;(\Varid{carrier}\;(\Varid{n}\mathbin{-}\mathrm{1}))~\boldsymbol{\mathop{\langle \star \rangle}}~\mathbf{Now}\;\Varid{p}\mathrel{\vee}{}\<[E]%
\\
\>[3]{}\hsindent{3}{}\<[6]%
\>[6]{}\mathbf{Now}\;(\Varid{carrier}\;(\Varid{n}\mathbin{+}\mathrm{1}))~\boldsymbol{\mathop{\langle \star \rangle}}~\mathbf{Now}\;\Varid{p}\;\mathbf{then}\;\mathrm{50}\;\mathbf{else}{}\<[E]%
\\
\>[B]{}\hsindent{3}{}\<[3]%
\>[3]{}\mathbf{if}\;\mathbf{Now}\;(\Varid{carrier}\;(\Varid{n}\mathbin{-}\mathrm{2}))~\boldsymbol{\mathop{\langle \star \rangle}}~\mathbf{Now}\;\Varid{p}\mathrel{\vee}{}\<[E]%
\\
\>[3]{}\hsindent{3}{}\<[6]%
\>[6]{}\mathbf{Now}\;(\Varid{carrier}\;(\Varid{n}\mathbin{+}\mathrm{2}))~\boldsymbol{\mathop{\langle \star \rangle}}~\mathbf{Now}\;\Varid{p}\;\mathbf{then}\;\mathrm{25}\;\mathbf{else}\;\mathrm{0}{}\<[E]%
\\
\>[B]{}\hsindent{3}{}\<[3]%
\>[3]{}\mathbf{where}{}\<[E]%
\\
\>[3]{}\hsindent{3}{}\<[6]%
\>[6]{}\Varid{carrier}\;\mathrm{0}\mathrel{=}\text{\ttfamily \char34 carrier\char95 smooth\char95 period\char34}\boldsymbol{<\!:}\mathrm{0}\boldsymbol{=:}(\equiv )~\boldsymbol{\mathop{\langle \$ \rangle}}~\mathbf{Now}\;\Varid{p}{}\<[E]%
\\
\>[3]{}\hsindent{3}{}\<[6]%
\>[6]{}\Varid{carrier}\;\Varid{n}\mathrel{=}\text{\ttfamily \char34 carrier\char95 smooth\char95 period\char34}\boldsymbol{<\!:}\Varid{n}\boldsymbol{=:}{}\<[E]%
\\
\>[6]{}\hsindent{3}{}\<[9]%
\>[9]{}\Varid{carrier}\;(\Varid{n}\mathbin{-}\mathrm{1})\boldsymbol{:\!\!@}(\mathbin{-}\mathrm{1},\mathbf{Leaf}\;(\Varid{const}\;\Conid{True})){}\<[E]%
\ColumnHook
\end{hscode}\resethooks
}

\noindent In the first experiment, we run all four specifications over traces
with synthetic inputs of varying length.
The results are shown in Fig.~\ref{tab:empirical:plots}~(a), which
suggest that the memory required is approximately constant,
indicating that the memory used is independent of the trace length,
and that monitors run in constant space, as theoretically predicted.

In the second experiment, we vary the period \ensuremath{\Varid{n}} to asses how the
number of streams affects the memory usage for both period checkers.
The outcome suggests that increasing the number of streams only
impacts linearly on the memory required to perform the monitoring, as
shown in Fig.~\ref{tab:empirical:plots}~(b).

In the third experiment, we use different values for the period \ensuremath{\Varid{n}} to
increase the absolute value of the \minBackRef for the Boolean and
quantitative period checkers to asses how increasing the absolute
value of the \minBackRef affects the memory required.
The outcome again suggests that the memory required grows linearly, as
shown in Fig.~\ref{tab:empirical:plots}~(c).
In both the second and third scenarios, we can observe that the memory
required is unaffected by whether we are working with quantitative
datatypes or \ensuremath{\Conid{Boolean}} values.
\section{Final Discussions, Conclusion and Future work}
\label{sec:conclusions}

\subsubsection{Final discussions}
One alternative to \ensuremath{\Conid{Typeable}} is to use modular datatypes and
evaluators~\cite{Swierstra:DTCarte}.
However, this would break our goal of transparently borrowing
datatypes in the lift deep embedding, by forcing \hlola data sorts to
be defined manually as Haskell datatypes.

Resource analysis is a central concern in RV and, in fact, in all
real-time and critical systems.
For example, aviation regulation forbids the use of runtime
environments with garbage collection for critical systems.
But this is still an option for soft-critical applications, where
\hlola has successfully been applied to improve mission software of
autonomous UAVs~\cite{zudaire20assumption}.
As future work we plan to generate embeddable C code from a restricted
version of \hlola, using the Ivory framework~\cite{IvoryLang} (see
Copilot~\cite{pike10copilot}).

An eDSL like \hlola is a library within the host language, and can be
used as a theory within \hlola reflectively.
This feature can greatly simplify writing specifications, used for
example to express predictive Kalman filters as
in~\cite{zudaire20assumption} or quantitative semantics of STL and
MTL.
\vspace{-1.2em}
\subsubsection{Conclusions}
We have presented \hlola, an engine for SRV implemented as a Haskell
eDSL.
We use the notion of lift deep embedding---folklore in advanced eDSLs
(see~\cite{westphal20implementing})---in a novel way to fulfill the
SRV promise of a clean separation between the temporal engine and the
data manipulated, allowing the transparent incorporation of new types.
Using Haskell makes readily available features like static
parameterization---which allows implementing many logics with Boolean
and quantitative semantics---, otherwise programmed in an ad-hoc
manner in other SRV tools.
The resulting system \hlola is very concise.
A well-known drawback of using an eDSL is that errors are usually
cryptic.
We are currently working on a front-end restriction of the language
that enables better error reporting, while still allowing expert users
to use all the advanced features.

Current work includes extending \hlola to support time-stamped event
streams, which allows monitoring real-time event sequences as
in~\cite{gorostiaga18striver}.
This extension will be to Striver~\cite{gorostiaga18striver} like
\hlola is to \lola.
From the point of view of exploiting Haskell further, future work
includes using LiquidHaskell more aggressively to prove properties of
specifications and memory bounds, as well as proving formally the
claim that our use of \ensuremath{\Conid{Dynamic}} is safe.
We are also working on using QuickCheck to generate test traces from
specifications and on studying how to use model-based testing to
improve the test suites obtained.
%

%%%%%%%%%%%%%%%%%%%
%
% APPENDIX
%
%%%%%%%%%%%%%%%%%%%

%\newpage

% \printbibliography
%\bibliographystyle{plain}
\bibliographystyle{abbrv}
\bibliography{references,orefs}

\end{document}